\documentclass[fleqn,usenatbib]{mnras}

\usepackage{newtxtext,newtxmath}

\usepackage[T1]{fontenc}

\DeclareRobustCommand{\VAN}[3]{#2}
\let\VANthebibliography\thebibliography
\def\thebibliography{\DeclareRobustCommand{\VAN}[3]{##3}\VANthebibliography}

\usepackage{graphicx}	
\usepackage{amsmath}	
\usepackage{setspace}
\usepackage{multirow}
\usepackage{subcaption}
\usepackage{type1cm}

\title[Cosmology with compound gravitational lenses]{Can we constrain cosmology with compound gravitational lenses?}

\author[D. J. Bowden et al.]{
Duncan J. Bowden,$^{1,2}$\thanks{E-mail: duncan.bowden@physics.ox.ac.uk}
Tian Li,$^{2}$
Aprajita Verma,$^{1}$
Thomas E. Collett$^{2}$
and Philip Holloway$^{2}$
\\
$^{1}$Department of Physics, University of Oxford, Keble Road, Oxford OX1 3RH, UK\\
$^{2}$Institute of Cosmology and Gravitation, University of Portsmouth, Burnaby Road, Portsmouth PO1 3FX, UK
}

\date{Accepted XXX. Received YYY; in original form ZZZ}

\pubyear{\the\year{}}

\begin{document}
\label{firstpage}
\pagerange{\pageref{firstpage}--\pageref{lastpage}}
\maketitle

\begin{abstract}
Compound lensing in double-source-plane lenses (DSPLs) introduces multiple mass-sheet transformations that bias the inferred distance ratios $\beta$ used to constrain cosmology. We present the recovered $\beta$ from lens models of simulated compound lenses and explore the systematic biases that arise from a mass-sheet transformation in the first source plane. We examined simulated observations based on SDSS J0946+1006, in which we varied the matter distribution of source 1. We found a strong correlation between the stellar-to-halo mass and the recovered $\beta$. We produced a simulated catalogue of DSPLs detectable by the Euclid telescope to see if the same systematic propagates to a more varied population. But at Euclid's resolution, the properties of source 1 were poorly constrained, and any correlated bias was lost in the statistical uncertainty. We inferred $w_0$ and $\Omega_{\rm m}$ for a flat $w$CDM cosmology from our $\beta$ posteriors for our combined sample. We found the cosmology biased when using the full mock Euclid sample, but reduced the bias substantially using a limited, well-modelled ``gold'' sub-sample, bringing it to within $\sim2\sigma$ and $\sim1\sigma$ of the ground-truth cosmology for simulated Euclid and Hubble Space Telescope imaging, respectively. Therefore, we must eliminate systems with unreliable lensing information, and for extreme compound lensing, the model complexity must be sufficient to reproduce the lensing observations. DSPLs remain a promising cosmological probe that could provide independent and complementary constraints to existing probes.
\end{abstract}




\section{Introduction}

Modern cosmological analyses push observations to the limit to test the widely accepted standard cosmological model: $\Lambda$ cold dark matter ($\Lambda$CDM). The cosmological constant $\Lambda$ assumes that dark energy is a static force that causes the accelerated expansion of the Universe. Many large surveys seek to determine whether a more complex model of dark energy is needed to describe the Universe, using well-established probes such as Type Ia supernovae \citep[SNe,][]{brout_pantheon_2022, abbott_dark_2024, rubin_union_2025}, the cosmic microwave background \citep[CMB,][]{planck_collaboration_planck_2020} and baryon acoustic oscillations \citep[BAO,][]{desi_collaboration_desi_2025}. While recent measurements indicate that dark energy may evolve with redshift \citep{desi_collaboration_desi_2025}, the significance of these results is not yet enough to rule out $\Lambda$CDM. In addition, SNe reanalyses have reduced the tension between observations and $\Lambda$CDM \citep[e.g.,][]{popovic_dark_2026}. Therefore, questions remain about whether systematic biases could produce this discrepancy. Although reducing uncertainty and bias from established probes will help, new independent measurements will be extremely beneficial in breaking degeneracies and validating existing constraints.

Strong gravitational lensing is one such phenomenon that can measure the evolution of the Universe across cosmic time. The variety of redshifts at which deflectors and sources can be found makes them appealing potential cosmological probes that can connect the low- and high-redshift Universe. Many unique classes of lenses can provide cosmological parameter constraints, most commonly the Hubble constant $H_0$ via time-delay cosmography of strongly lensed quasars and supernovae \citep{collaboration_tdcosmo_2026, pierel_cosmology_2026}. Here, we will focus on galaxy-scale double-source-plane lenses (DSPLs). DSPLs are gravitational lensing systems with two strongly lensed galaxies at distinct redshifts, source 1 (S1) and source 2 (S2). This configuration allows for compound lensing in which the light from the most distant source is deflected by the lens and S1. The Einstein radii ratio of the two lensed sources is related to a ratio of the angular diameter distances pertaining to the lens and sources. This distance ratio makes DSPLs sensitive to the matter density parameter $\Omega_{\rm m}$ and the dark energy equation of state parameters (e.g., $w_0, w_a$) \citep{collett_constraining_2012}. Therefore, using redshift information and measuring the distance ratio through detailed lens modelling, we can infer these cosmological parameters \citep[e.g.,][]{collett_cosmological_2014, sahu_cosmography_2025, bowden_constraining_2025}. 

With the abundance of strong lenses now being discovered by Euclid \citep{collaboration_euclid_2025}, and many more to come from the complete Euclid Wide Survey (EWS), the Vera C. Rubin Observatory's Legacy Survey of Space and Time (LSST) and the Nancy Grace Roman Space Telescope's High-Latitude Wide-Area Survey, we will have access to $\mathcal{O}(10^5)$ strong lensing systems \citep{collett_population_2015, holloway_detectability_2023, ferrami_model_2024, shajib_strong_2025}. The improvement for DSPL is especially significant since we will transition from a handful of known systems to $>10^3$ candidates from Euclid \citep{collett_population_2015, li_euclid_2025}, giving us the advantage of large-number statistics previously unavailable. Spectroscopic surveys will provide the information required to confidently determine the redshifts of lens and source galaxies, either serendipitously in wide-area surveys such as the DESI Legacy Imaging Surveys \citep{dey_overview_2019} or through targeted strong lensing follow-up such as the 4MOST Strong Lensing Spectroscopic Legacy Survey \citep[4SLSLS,][]{collett_4most_2023} which aims to obtain $\sim 5000$ source and deflector redshifts. With this wealth of observations, strong lenses are set to become a promising independent probe for cosmology \citep{sharma_testing_2023, shajib_strong_2025}. 

As for any cosmological probe, we must identify and mitigate any systematic biases in the constraints inferred from DSPLs. A well-known source of bias in any strong lensing system is the mass-sheet degeneracy \citep{falco_model-dependent_1985, schneider_mass-sheet_2013}. A mass-sheet transformation (MST) describes the invariant transformation between the angular size of a source and the lensing mass of the deflector, i.e., the same lensed images can be produced by a smaller source given the mass of the lens is greater or vice versa. From lensing information alone, it can be challenging to find the most appropriate choice from this family of solutions. Typically, the MST can be split into internal and external components \citep{koopmans_gravitational_2004, suyu_dissecting_2010, birrer_tdcosmo_2020}. For DSPLs, the freedom of the MST internal to the first deflector (hereafter, the lens) is reduced by the presence of two sources with images that probe different angular scales. However, there will always be some mismatch between the true mass distribution and our assumed profile. The external component of the MST comes from line-of-sight (LOS) perturbers in the foreground of the sources and deflectors \citep{birrer_tdcosmo_2020, fleury_line--sight_2021}. Compound lensing introduces a second MST within the first source plane \citep{schneider_can_2014}, which is often ignored and has not been studied in detail. These mass-sheet degeneracies can strongly influence the cosmological parameters that can be inferred from these systems \citep{schneider_can_2014, schneider_generalized_2019}. With the addition of stellar kinematic measurements and galaxy number counts, several studies have shown that the MST can be constrained and explicitly marginalised, with the TDCOSMO collaboration in particular doing this for a sample of lensed quasars with precise cosmology in mind \citep[e.g.,][]{collaboration_tdcosmo_2026, sheu_tdcosmo_2026}. Although kinematic modelling of lensed sources is possible with spatially resolved kinematics \citep[e.g.,][]{newman_resolving_2018, g_c_spatially_2025}, it has never been used to constrain the MST in compound lenses. For that reason, in this paper we use end-to-end simulations of compound lenses to explore the bias associated with the first source plane in a Bayesian framework.  

We investigated a family of simulated systems based on the first discovered DSPL SDSS J0946+1006 \citep{gavazzi_sloan_2008}, henceforth referred to as the Jackpot. The Jackpot has a tight LOS alignment, causing both source galaxies to be quadruply imaged and allowing for a well-constrained lens model. From this, we simulated observations while varying the mass distribution of the intermediate source S1 using a stellar + dark matter model. We produced lens models of our simulations using a simple mass parameterisation for S1 and examined how the MST between the true and chosen mass profile biased the inferred distance ratio. Next, we simulated a sample of compound lenses identifiable in the EWS by constructing simulated observations and using detectability constraints to generate a population of systems. We then produced lens models to test how any bias introduced by S1 propagates into a more varied sample of compound lenses. In addition, we investigate how complementary high-resolution (0.05\arcsec) HST follow-up of the ``gold'' lenses (Section \ref{sec:gold}) from our sample alters our inferred cosmology.

In this paper, we summarise compound gravitational lensing theory for double-source-plane systems in Section \ref{sec:theory}. In Section \ref{sec:sims}, we outline how we produced compound lens simulations and their mock observations. These simulations are divided into the aforementioned Jackpot and Euclid categories and are constructed as described in Sections \ref{sec:jackpot-like} and \ref{sec:euclid-pop}. Following this, we detail the methodology used to produce lens models of our mocks in Section \ref{sec:modelling} and present the results of these models and the cosmological constraints inferred from them in Section \ref{sec:results}. We discuss the implications of our findings in Section \ref{sec:discussion}, provide recommendations for future cosmological DSPL studies and finally summarise these findings in Section \ref{sec:conclusion}. Throughout the simulations presented in this work, we use a consistent cosmological model of a flat $\Lambda$CDM cosmology with a matter density fraction of $\Omega_{\rm m} = 0.3$ and Hubble constant $H_0=70\rm{kms^{-1}Mpc^{-1}}$ where applicable.

\section{Compound Gravitational Lensing}\label{sec:theory}

Compound gravitational lensing occurs in multi-source-plane strong lenses where two or more deflector galaxies are located along the LOS in different redshift planes. In this scenario, the lensed galaxies can themselves be lenses, adding to the deflection of more distant source galaxies. This multi-plane deflection is known as compound lensing, schematics of which can be found in Figure 1 of \cite{bowden_constraining_2025}. The lens equation, which describes the deflection of the true source position to the observed position, can be generalised to compound lenses with any number of deflector planes \citep{schneider_generalized_2019}. The unlensed angular position of a source in the $j^{th}$ redshift plane $\theta_j$ can be expressed as a function of the observed angular position in the $i^{th}$ redshift plane $\theta_i$ by:

\begin{equation}\label{eq:multi}
    \theta_{j} = \theta_{1} - \sum_{i=1}^{j-1}\beta_{ij}\alpha_i^{\prime}(\theta_i),
\end{equation}
where $\theta_1$ is the observed angular position of the source in the image plane ($1^{st}$ deflector plane), and $\alpha_i^{\prime}$ is the scaled deflection angle of the $i^{th}$ deflector plane, a factor of $\frac{D_{is}}{D_s}$ times the physical deflection angle $\hat{\alpha}_i$. The distance ratio $\beta_{ij}$ is given by:
\begin{equation}
    \beta_{ij} = \frac{D_{ij}D_s}{D_jD_{is}},
\end{equation}
where $D$ is an angular diameter distance between the observer and $i^{th}$ redshift plane or between redshift planes $i,j$, with $s$ denoting the redshift plane of the most distant source.

In this work, we will be considering the simplest compound lensing case where there are two source planes. Therefore, we can simplify Equation \ref{eq:multi} to write the double-source-plane lens equation:
\begin{equation}\label{eq:theta3}
    \theta_{3} = \theta_{1} - \alpha_1^{\prime}(\theta_1) - \alpha_2^{\prime}(\theta_2),
\end{equation}
which describes the unlensed angular position of S2 $\theta_3$ as a function of its observed position $\theta_1$, the scaled deflection angle of the lens $\alpha_1^{\prime}$, the scaled deflection angle of S1 $\alpha_2^{\prime}$ and the angular position in the first source plane $\theta_2$ given by:
\begin{equation}\label{eq:theta2}
    \theta_2 = \theta_1 - \beta_{12}\alpha_1^{\prime}(\theta_1),
\end{equation}
where $\beta_{12}$ is the distance ratio;
\begin{equation}\label{eq:beta}
    \beta_{12} = \frac{D_{12}D_3}{D_2D_{13}},
\end{equation}
henceforth referred to as $\beta$. All angles and distances are defined as in Figure 1 of \cite{bowden_constraining_2025}.

By building mass models that produce the required deflection angles $\alpha_i^{\prime}$, one can simultaneously solve for $\beta$ and the remaining lens-equation parameters if given the observed angular positions $\theta_1$. The details of this lens modelling are given in Section \ref{sec:modelling}.

\subsection{Cosmological Inference}

Within a flat $w$CDM cosmology, the angular diameter distance between redshifts $z_i$ and $z_j$ along the LOS is given by:
\begin{equation}\label{eq:dist}
    D_{ij}=\frac{c}{H_{0} (1+z_j)} \int_{z_i}^{z_{j}} \frac{dz}{E(z)}, 
\end{equation}
where $c$ is the speed of light, $H_0$ is the Hubble constant and the normalized Hubble parameter $E(z) \equiv H(z)/H_0$ can be expressed as:
\begin{equation}\label{eq:E(z)}
    E(z)=\sqrt{ \Omega_{\rm m} (1+z)^3 +  (1-\Omega_{\rm m})(1+z)^{3(1+w)} },
\end{equation}
where $\Omega_{\rm m}$ is the matter density parameter and $w_0$ is the dark energy equation of state parameter. Consequently, $\beta$ is a function of only redshift, $w_0$ and $\Omega_{\rm m}$ due to the factor of $H_0$ cancelling. Therefore, we can place constraints on $w_0$ and $\Omega_{\rm m}$ through lens model inference of $\beta$ and redshift measurements. The same principle applies to alternative cosmological models, given they have an analytical definition for $E(z)$.

\subsection{Mass-sheet transformation}

There is a major caveat that dominates systematic uncertainty in strong lensing analyses \citep[e.g.,][]{gomer_galaxy-lens_2020, oriordan_angular_2024, collaboration_tdcosmo_2026}. For any given lens, there is a mathematical transformation that allows the observed angular positions of the source to remain unchanged while the mass of the lens and distance to the source can vary; this is known as the mass-sheet transformation (MST). Increasing the mass of the lens causes an increase in deflection, but moving the source closer to the lens counteracts the additional deflection and vice versa. By adding uniform mass sheets to the lens, we can achieve the same observables with the correct scaling. This propagates to compound lenses where multiple deflector planes allow multiple mass sheets.

Equations \ref{eq:theta3} and \ref{eq:theta2} remain unchanged when an MST is applied, but the parameters measured from modelling are subject to the MST's scaling. Following \cite{schneider_can_2014}, we can introduce the transformation into the unlensed source positions;
\begin{equation}
    \tilde\theta_{3} = \theta_{1} - \tilde{\alpha}_1(\theta_1) - \tilde{\alpha}_2(\tilde\theta_2) = \nu_3\theta_3,
\end{equation}
and
\begin{equation}
    \tilde\theta_2 = \theta_1 - \tilde\beta\tilde\alpha_1(\theta_1) = \nu_2\theta_2,
\end{equation}
where tildes represent the transformed quantities, $\tilde\alpha_1$ and $\tilde\alpha_2$ are defined in \cite{schneider_can_2014}. $\nu_2$ and $\nu_3$ represent the source position scaling in the 2nd and 3rd redshift planes and are defined as:
\begin{equation}\label{eq:nu}
    \nu_2 = \frac{\tilde\beta\lambda}{\beta}, \ \ \ \nu_3 = \frac{\beta}{\tilde\beta}\frac{1-\tilde\beta}{1-\beta}.
\end{equation}
This choice leaves us with mass sheets $K_1$ and $K_2$ in the first and second deflector planes, respectively;
\begin{equation}\label{eq:K}
    K_1 = \frac{1}{\tilde\beta} - \frac{\lambda}\beta{}, \ \ \ K_2 = \frac{1}{\tilde\beta}\left(1-\frac{\beta}{\lambda\tilde\beta}\frac{1-\tilde\beta}{1-\beta}\right).
\end{equation}
This allows us to rescale our unknown source positions in each source plane and the deflection angles in each deflector plane in addition to our distance ratio, while preserving the observed position by introducing a single mass-sheet transformation parameter $\lambda$. Therefore, lens models that do not explicitly parameterise and constrain the MST will infer the rescaled distance ratio $\tilde{\beta}$.

The MST alone is mathematical, but we can attribute physical meaning to this transformation driven by external and internal factors within a galactic mass profile. An external MST describes the uniform convergence field which can arise due to relative overdensities or underdensities in the local environment and LOS based on the number of objects within them. An internal MST describes the mismatch between a chosen mass profile parameterisation and a galaxy's true mass distribution. We can mitigate how much degeneracy is allowed through estimates of the local environment density \citep[e.g.,][]{greene_improving_2013, wells_tdcosmo_2023} and independent measurements of the galaxy's mass profile \citep{tdcosmo_collaboration_tdcosmo_2025}. However, the differences between a true galaxy and commonly used parameterisations leave room for the MST.

In this work, we explored the bias that deflection by S1 can introduce when the ground-truth mass distribution is more complex than our model. The difference between the true and model profiles allows for an internal MST that modifies the Einstein radius and density slope of a mass profile when not explicitly parameterised. We assumed that the mass profile of the lens can be well constrained with the addition of independent information such as stellar kinematics and therefore do not introduce room for an internal MST. Therefore, we set $K_1 = 0$ resulting in $\lambda=\beta/\tilde\beta$ from Equation \ref{eq:K} and $\nu_2=1$ in Equation \ref{eq:nu}.

\section{Simulated Observations}\label{sec:sims}

We constructed simulated DSPL images using the open-source strong lensing codes \textsc{Herculens}\footnote[1]{\url{https://github.com/Herculens}} \citep{galan_using_2022, galan_gordo_2024, enzi_overconcentrated_2025} and \textsc{Lenstronomy} \footnote[2]{\url{https://github.com/lenstronomy}}\citep{birrer_lenstronomy_2018, birrer_lenstronomy_2021} to simulate Jackpot-like and Euclid observations. Each simulated DSPL is structured as three redshift planes corresponding to the lens, S1 and S2 from lower to higher redshift. Each plane has an associated light component, and the lens and first source planes have a mass component. Within this framework, we can appropriately account for the compound lensing by the lens and S1. The choice of light and mass profiles used for each component is described in Sections \ref{sec:jackpot-like} and \ref{sec:euclid-pop}, as these differ for the Jackpot-like and Euclid population of DSPLs.
 
\subsection{Jackpot-like}\label{sec:jackpot-like}

In the case of the Jackpot, we take a simple approach, starting from a lens model of the real Jackpot, using a setup similar to \cite{collett_cosmological_2014}, and interchanging the mass model for S1 between mock observations. Our simulated HST observations mimic those of \cite{gavazzi_sloan_2008}. We matched the properties of HST's Advanced Camera for Surveys (ACS) F814W filter, using four exposures with a total exposure time of 2096 seconds and a $0.05\arcsec$ pixel scale. We matched the estimated root-mean-square (RMS) of the background noise to that of the HST observations, which is equivalent to $\sim 0.015$ counts per second.

We omit the lens light from our simulated DSPLs as this is inferred independently of the lens mass, and its exclusion reduces the modelling computation time. The light of S1 and S2 is described by pixelated maps of surface brightness taken from the posterior median model of the Jackpot. The mass of the lens is described by an elliptical power law (EPL) profile \citep{tessore_elliptical_2015}, plus shear to account for external shear and residual complexity in the primary lens profile, as is done by \cite{collett_cosmological_2014}. The values of the ground-truth mass-model parameters and distance ratio $\beta$ that remain unchanged in each simulation are listed in Table \ref{tab:truth}.

\begin{table}
    \centering
    \begin{tabular}{|r|llllll|}
        \textbf{Component} & \multicolumn{6}{l}{\textbf{Parameter Values}} \\
        \hline
        EPL (lens) & $\theta_{\rm E}$ & $\gamma$ & $e_1$ & $e_2$ & $x_0$ & $y_0$ \\
         & 1.398 & 2.032 & 0.003 & -0.029 & 0.022 & 0.047 \\
        \hline
        Shear (lens) & $\gamma_1$ & $\gamma_2$ &  &  &  & \\
         & 0.049 & -0.058 & &  &  & \\
        \hline
        Distance ratio & $\beta$ \\
         & 0.730
    \end{tabular}
    \caption{Ground-truth model parameters used as the base to produce mock Jackpot variations. Values are taken from the median of the Jackpot lens model posteriors.}
    \label{tab:truth}
\end{table}

The mass profile of S1 is not derived from the Jackpot, and instead replacements are drawn from the CosmoDC2 synthetic galaxy catalogue \citep{korytov_cosmodc2_2019}. CosmoDC2 is derived from an N-body simulation covering 440 degrees$^2$ of the sky up to $z=3$ and contains the physical properties of galaxies and the dark matter halos in which they reside. To inject the galaxies from CosmoDC2 into the Jackpot model, we needed to convert the physical properties into lens model parameters for our chosen mass profiles. We assumed that the total mass of each galaxy is composed of a stellar component, described by a Sérsic profile \citep{sersic1968atlas}, plus a dark matter component described by a Navarro-Frenk-White (NFW) profile \citep{navarro_universal_1997}. We used the elliptical cuspy NFW and elliptical Sérsic convergence parameterisations defined by \cite{keeton_catalog_2002}. 

\begin{figure}
    \centering
    \includegraphics[width=1\linewidth]{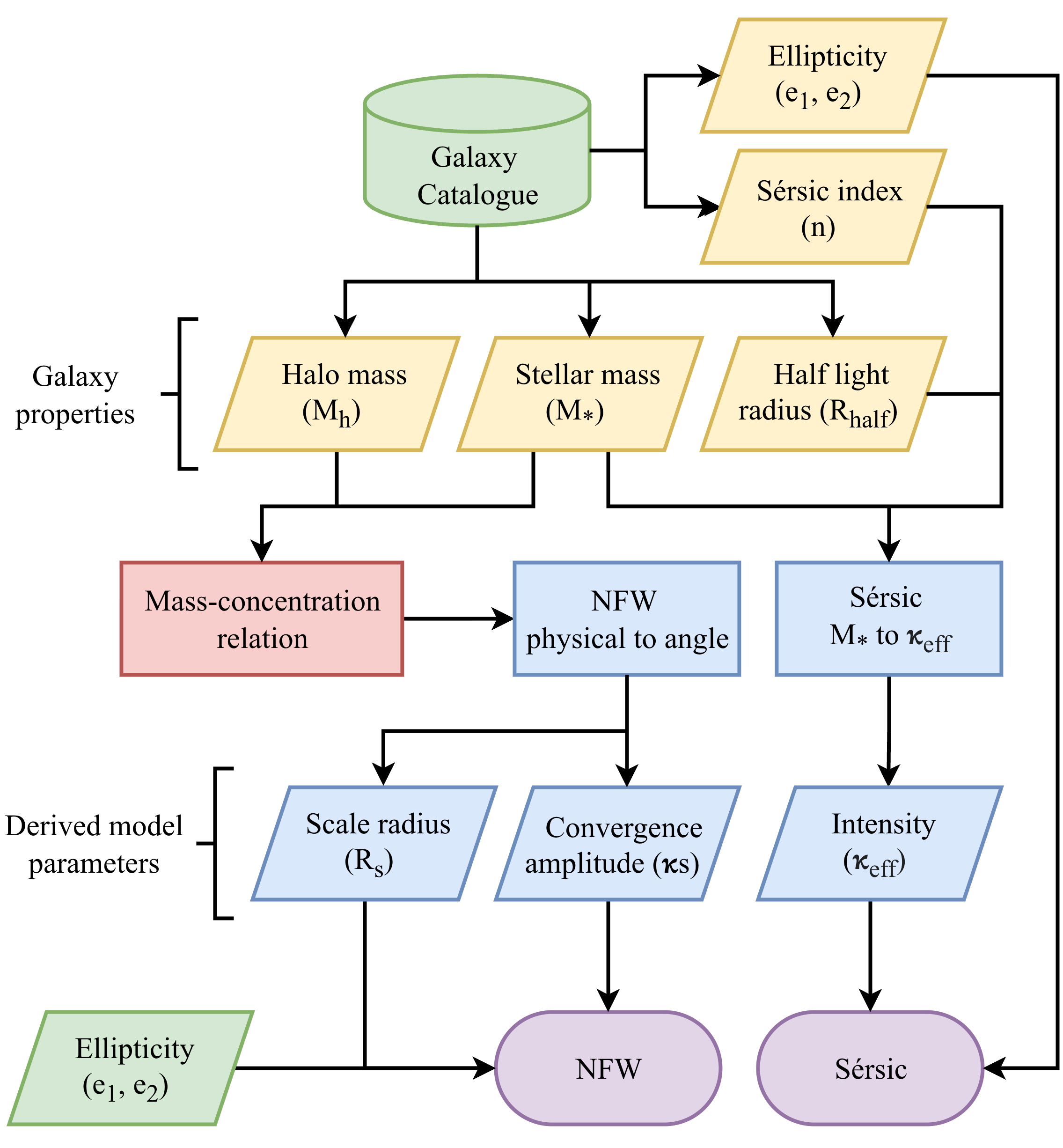}
    \caption{Schematic diagram for the conversion of galaxy properties into lensing mass-model parameters for the chosen NFW and Sérsic profiles for source 1. Parallelograms represent inputs and outputs, rectangles are processes and cylinders are catalogues/databases. The mass-concentration relation is given by Equation \ref{eq:mass-conc}. The NFW physical to angle and Sérsic $M_*$ to $\kappa_{\rm{eff}}$ processes are performed through \textsc{Lenstronomy}.}
    \label{fig:mass_param}
\end{figure}

The process by which we convert the galaxy properties from the catalogue into mass-model parameters is depicted in Figure \ref{fig:mass_param} and described in the following paragraph. For the Sérsic, we derived the intensity $\kappa_{\rm{eff}}$ from the stellar mass $M_*$, half-light radius $R_{\rm{half}}$ and Sérsic index $n$ within \textsc{Lenstronomy}. The remaining Sérsic parameters match one-to-one with $R_{\rm{half}}$, $n$ and the ellipticity parameters $e_1$ and $e_2$ from the catalogue. For the NFW, we derived the scale radius $R_{\rm{s}}$ and convergence amplitude $\kappa_{\rm{s}}$ from the halo mass $M_{\rm{h}}$ and the concentration parameter $c$, using \textsc{Lenstronomy}. We computed $c$ using the concentration-mass relation, given by Equation 18 from \cite{child_halo_2018}:
\begin{equation}\label{eq:mass-conc}
    c = A((\frac{M_{\rm{h}}}{M_{*}b})^m(\frac{M_{\rm{h}}}{M_{*}(b+1)})^{-m}-1)+c_0,
\end{equation}
where $A, m, b$ and $c_0$ are given in Table 1 of \cite{child_halo_2018} for ``Individual, all''.

\begin{figure}
    \centering
    \includegraphics[width=\linewidth]{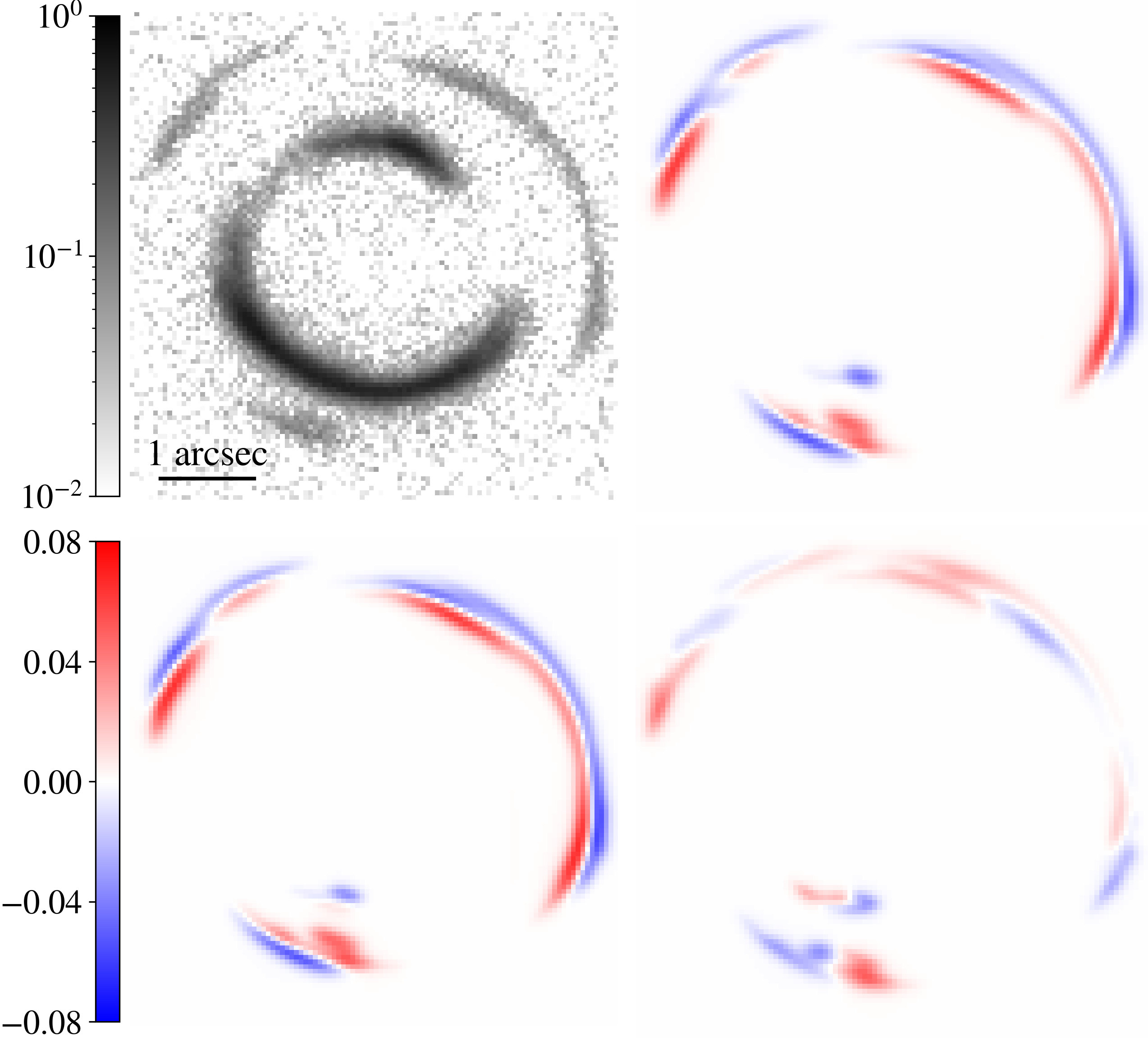}
    \caption{Top left, an example mock observation of the Jackpot with background and Poisson noise but excluding lens light. Remaining panels: difference images between the example and three other mock Jackpots, excluding any noise.}
    \label{fig:example}
\end{figure}

The centre of mass $(x_0, y_0)$ of S1 is taken from the Jackpot posteriors, which is aligned with the visible counterpart within the light model and remains unchanged throughout the simulations. The NFW and Sérsic profiles share the same S1 centre, but we do not assume that the ellipticity of the NFW follows that of the Sérsic and instead derive it from a uniform distribution of axis ratios $q\in [0.8, 1]$ since dark matter halos are close to spherical \citep{schrabback_tightening_2021} and we do not assume that their axes are aligned with the galaxy. From these parameters, simulated observations are reconstructed using \textsc{Herculens}. Figure \ref{fig:example} includes an example mock observation and difference images between that example and three other simulations. We can see that only the lensed images of S2 vary because only the deflection by S1 changes. However, we do see a small amount of overlapping light due to compound lensing, producing a faint image of S2 at the centre of S1's arc.

\subsubsection{Sample selection}

\begin{figure*}
    \centering
    \includegraphics[width=1\linewidth]{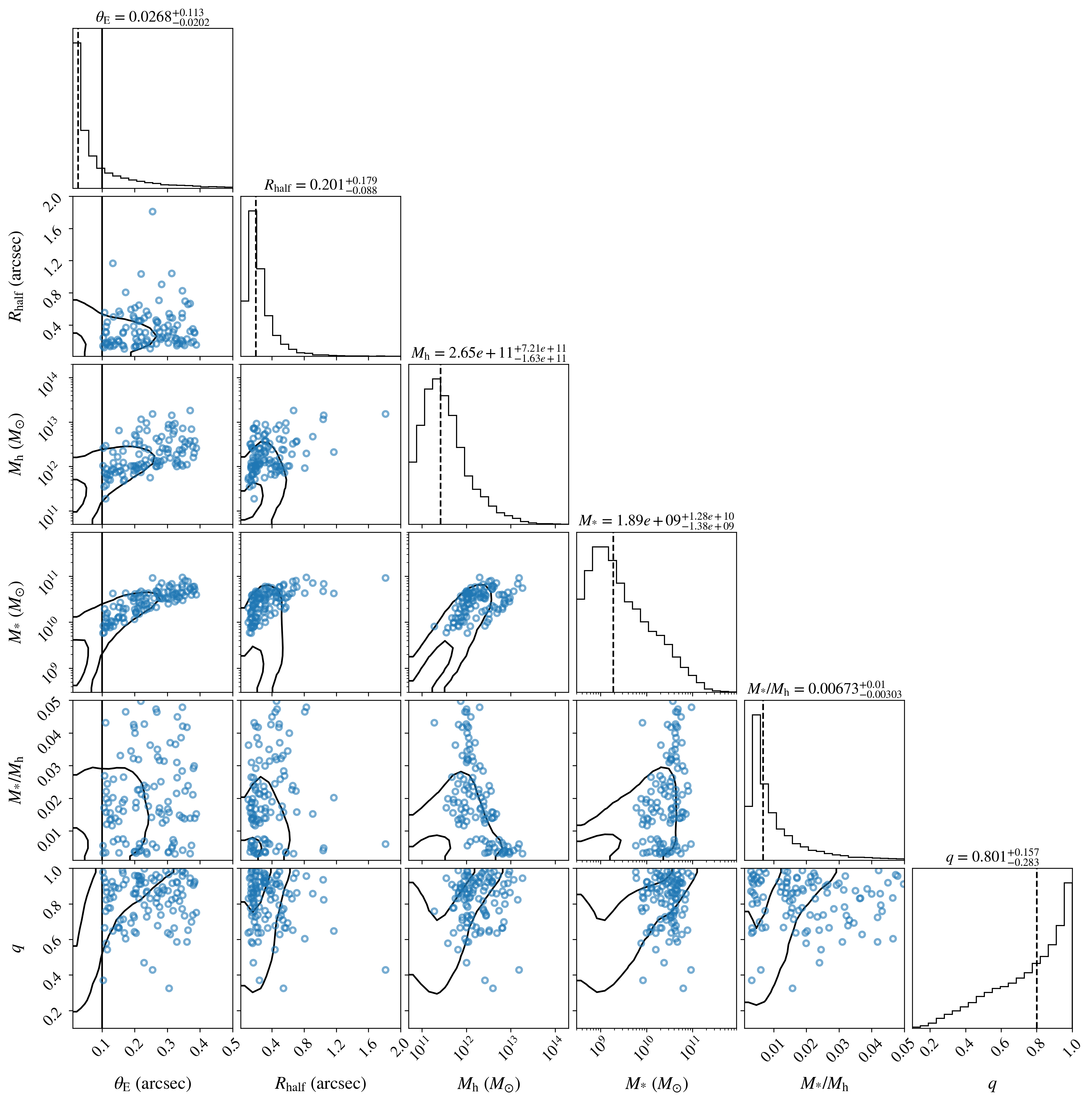}
    \caption{Two-dimensional distributions of the effective Einstein radius $\theta_{\rm E}$, half-light radius $R_{\rm{half}}$, halo mass $M_{\rm h}$, stellar mass $M_*$, stellar-to-halo mass ratio $M_{*}/M_{\rm h}$ and axis ratio $q$ of galaxies sampled from the CosmoDC2 galaxy catalogue. The black contours represent the distribution of galaxy properties within the catalogue, bounding the regions within which 39.3\% and 86.4\% of the galaxies lie, and the blue circles highlight the 110 galaxies drawn for our sample. Panels along the diagonal show the one-dimensional distribution of each property with their median values indicated by the dashed vertical lines. The solid vertical line signifies the lower bound placed on $\theta_{\rm E}$ at 0.1.}
    \label{fig:sample_corner}
\end{figure*}

We take a random subset of $\sim10,000$ galaxies from the CosmoDC2 catalogue. We calculated the effective $\theta_{\rm E}$ by numerically solving for the angular radius within which the radial convergence profile of our NFW + Sérsic profile integrates to 1. To ensure that all of our simulations produce some amount of observable compound lensing, we sampled galaxies uniformly with effective Einstein radii $\theta_{\rm E} \in [0.1, 0.4]$. In addition, we restricted the halo-to-stellar mass ratio of our sample to be $\in [20, 350]$, to select galaxy-scale systems consistent with predictions and observations of the galactic stellar-to-halo mass relation \citep{girelli_stellar--halo_2020, shuntov_cosmos2020_2022}. The distribution of key galaxy properties within the sample is shown in Figure \ref{fig:sample_corner}. We randomly sampled 110 galaxies and used them to produce 110 variations of the simulated Jackpot-like observations; the modelling of these is later described in Section \ref{sec:modelling}.

\subsection{Euclid population}\label{sec:euclid-pop}

For generating the more general Euclid simulations, we take a more holistic approach starting with a simulated galaxy catalogue and building lenses from scratch. For our population of Euclid compound lenses, we simulated Euclid observations, and for a subset of ``gold'' lenses we also simulated HST imaging, using the same observational properties as described for the Jackpot simulations. The gold sample of lenses is designed to emulate systems discovered by Euclid and then selected for follow-up with HST; further details of this sample can be found in Section \ref{sec:gold}. 

We simulated our Euclid observations to emulate the VIS filter with a resolution of $0.1\arcsec$. We match the properties of the EWS, using four exposures with a total exposure time of 2242 seconds and read noise of 4.5 counts per exposure \citep{euclid_collaboration_euclid_2025}, which dominates the background RMS. Euclid was chosen due to being a major pathway for current and future DSPL discovery \citep{collett_population_2015, li_euclid_2025}. The EWS will cover $\sim14000\ \rm{deg}^2$ of the sky once complete, and will soon release observations for $\sim1900\ \rm{deg}^2$ from its data release 1 (DR1) (\hyperlink{https://www.cosmos.esa.int/web/euclid/euclid-dr1}{Euclid Collaboration in prep.}).

Unlike the Jackpot-like simulations, the properties for each mass and light component in our mock Euclid population are drawn from a galaxy catalogue. For the lens light, we include the Poisson noise associated with an elliptical Sérsic light profile and a Sérsic index of 4 but omit the light itself, assuming a Poisson-limited lens-subtracted image. The two sources are also represented using elliptical Sérsic profiles with a fixed Sérsic index of 2. The mass of the lens is described using a singular isothermal ellipsoid (SIE) \citep{kormann_isothermal_1994} plus shear drawn from $\mathcal{U}[-0.2, 0.2]$. An SIE + shear was chosen as a relatively simple yet commonly used model for our primary lens, where the shear represents both external shear and additional complexity within the galaxy mass profile. We used the same elliptical NFW + elliptical Sérsic parameterisation to describe the mass profile of S1 as for the Jackpot-like simulations, and derived the mass profile parameters in the same manner (Section \ref{sec:jackpot-like}).

\begin{figure}
    \centering
    \includegraphics[width=1\linewidth]{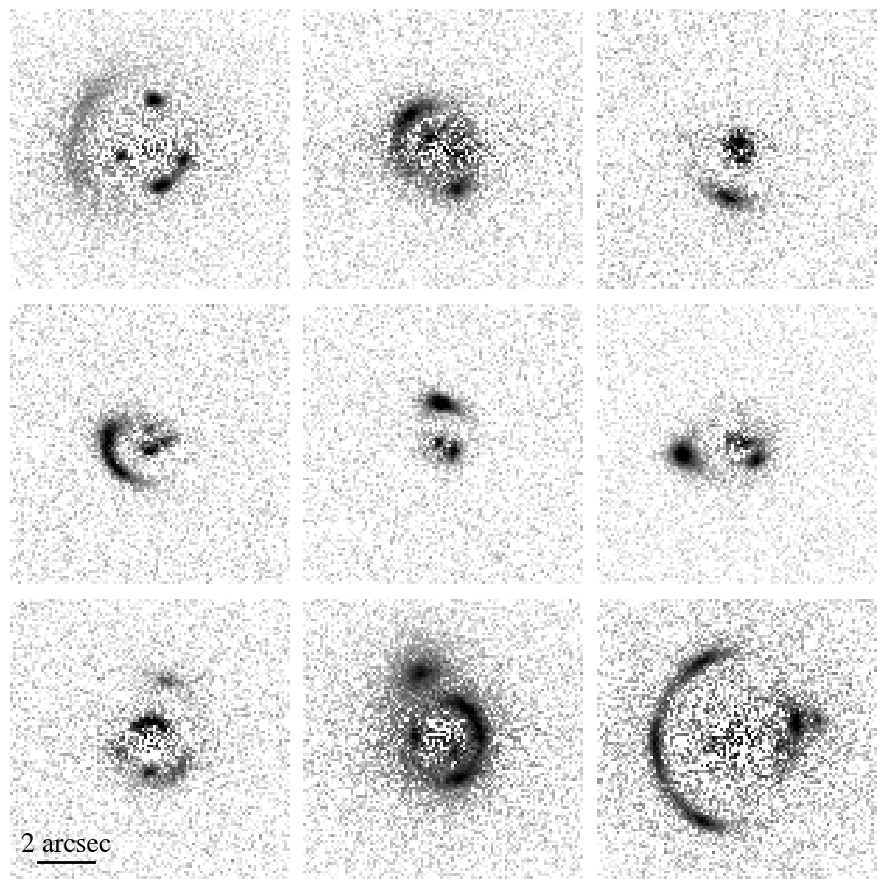}
    \caption{Simulated Euclid observations of 9 compound lenses determined to be detectable by the Euclid Wide Survey. Lens light is omitted with the assumption of Poisson-limited lens subtraction.}
    \label{fig:euclid_sims}
\end{figure}

We used \textsc{LensPop}\footnote[3]{\url{https://github.com/tcollett/LensPop}} \citep{collett_population_2015} to compile the DSPL catalogue, simulate lens properties and enforce criteria, detailed in Section \ref{sec:criteria}, to produce a population across $4200\ \rm{deg}^2$ of the sky observable by the EWS. By default, \textsc{LensPop} places signal-to-noise ratio (SNR) constraints on S1 and S2 using the total integrated light from the images of each source. However, we remove this constraint and perform our own SNR computation assuming a perfect lens light subtraction but including Poisson noise from the lens, as described in Section \ref{sec:SNR}. We opted for this to maintain control over the limiting SNR on a per-image basis. Because we intended to perform lens model inference on the selected DSPLs, we had to ensure that each source had at least one counter image present with a large enough (detectable) signal for modelling. Examples of such detectable compound lens arcs with simulated Euclid imaging are shown in Figure \ref{fig:euclid_sims}.

\subsubsection{Selection and detectability criteria}\label{sec:criteria}

Following a similar procedure to \cite{holloway_detectability_2023}, we performed selection cuts based on the lensing properties of the lens, S1, and S2 in our DSPL catalogue. The criteria are as follows:
\begin{spacing}{1.5}
\begin{enumerate}
    \item $z_{\rm s} > z_{\rm l}$,
    \item $\theta_{\rm s} < \theta_{\rm E}^{\rm s}$ (i.e. the angular position of the unlensed source is within the Einstein radius),
    \item $\theta_{\rm E}^{\rm s1} > 0.1$
    \item $\theta_{\rm E}^{\rm s2} > 0.5$,
    \item $\mu_{\rm s} > 3$,
    \item $\mu_{\rm s} < 100$ (preventing point-like sources close to the caustic from producing unphysical magnifications), 
    \item $\mu_{\rm s} \cdot R_{\rm s} > 0.3$ (i.e. the effective size of lensed images is greater than 0.3),
    \item $\beta < 0.95$ (preventing sources from being too close in distance from the observer),
\end{enumerate}
\end{spacing}
where $z_{\rm s}$ is the redshift of a source and $z_{\rm l}$ is the redshift of the lens. $\theta_{\rm s}$ is the angular separation between the lens and the unlensed position of a source, and $\theta_{\rm E}^{\rm s}$ is the Einstein radius of the lens with respect to a source with redshifts $z_{\rm l}$ and $z_{\rm s}$. $\theta_{\rm E}^{\rm s1}$ and $\theta_{\rm E}^{\rm s2}$ are the Einstein radii of the lens with respect to S1 and S2. $\mu_{\rm s}$ is the source magnification, $R_{\rm s}$ is the half-light radius of the unlensed source and $\beta$ is the cosmological distance ratio. These criteria were applied with respect to each source, and the sources in each given DSPL were assigned labels $\rm s1$ and $\rm s2$ so that $z_{\rm s1} < z_{\rm s2}$.

We now define what we refer to as a compound lens throughout the rest of this work. $\theta_{\rm E,s1}$ is the effective Einstein radius of S1, defined as the radius within which the mean convergence is 1. We require $\theta_{\rm E,s1} > 0.1$ for a DSPL to be classified as a compound lens. This definition does not encompass all compound lenses since any DSPL with a non-zero $\theta_{\rm E,s1}$ can have a compound lensing effect. However, it is mostly perturbative below our chosen threshold of 0.1. In addition, not all systems with $\theta_{\rm E,s1} > 0.1$ necessarily have a detectable impact from compound lensing due to poor source alignment.

\subsubsection{Signal-to-noise ratio}\label{sec:SNR}

For each of the compound lenses in the catalogue, we generated simulated SNR maps for S1 and S2 and assumed Poisson-limited lens light subtraction for arc discovery. We computed the incident photons per pixel for each source in accordance with the total number of photons collected by the Euclid telescope:
\begin{equation}
    n = 10^{\frac{ZP-m}{2.5}}\cdot t_{\rm exp},
\end{equation}
for an object with AB magnitude $m$, observed through a filter with an AB magnitude zeropoint $ZP$ for an exposure time of $t_{\rm exp}$. The noise per pixel for each source was calculated as the sum in quadrature of background noise $\sigma_{\rm b}$ (dominated by read noise), Poisson noise from the lens $\sqrt{n_{\rm l}}$ and sources $\sqrt{n_{\rm s}}$, and the signal of the other source $n_{\rm s}^\prime$, given as follows:
\begin{equation}
    N_{\rm s} = \sqrt{\sigma_{\rm b}^2 + (\sqrt{n_{\rm l}})^2 + (\sqrt{n_{\rm s}})^2 + n_{\rm s}^{\prime 2}}.
\end{equation}
We included the $n_{\rm s}^\prime$ term to account for the overlap between images of the two lensed sources, which can obscure whether the sources can be distinguished from one another. To associate an SNR for each lensed image, we produced an image segmentation map that associates every region within the source SNR map with a peak corresponding to each resolvable image. We performed a gradient ascent on the SNR map of each source and set an SNR threshold to exclude all pixels with SNR $<1$. To compute the SNR of each image, we sum over each connected pixel $i$ within the image segment:
\begin{equation}
    \rm{SNR} = \frac{\Sigma_iS_i}{\sqrt{\Sigma_iN_i^2}}.
\end{equation}
This step allowed us to place constraints on the minimum SNR of images to deem each lensed source detectable. We required that for each source, at least one image has an SNR $\geq 10$ and that there is a second image present with an SNR $\geq 7$ for the compound lens to be detectable. We chose these limits to enable reliable lens modelling, but there may be more DSPLs identifiable by eye from the simulated VIS imaging.

\begin{figure}
    \centering
    \includegraphics[width=1\linewidth]{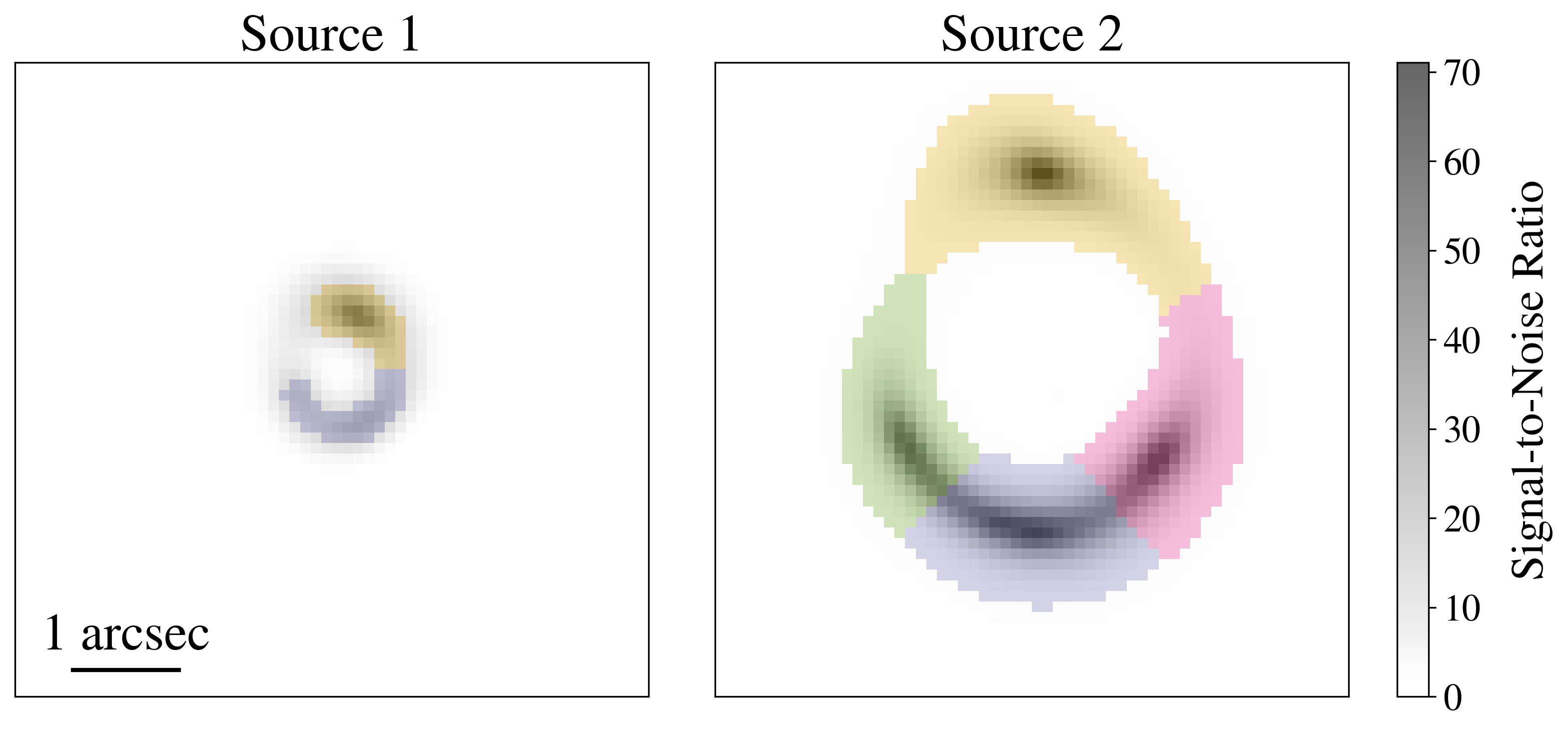}
    \caption{Example segmentation maps for the lensed images of sources 1 and 2. Each segment is shaded in a different colour and is associated with a signal-to-noise peak representing the centre of a lensed image. Each pixel is shaded by its individual signal-to-noise ratio.}
    \label{fig:SNR}
\end{figure}

In addition to the SNR constraints, we placed restrictions on the percentage of overlap in the area between the images from the two sources. This is in part to exclude systems that have images with little separation between sources, which would not be identifiable as DSPLs. More importantly, we did this to eliminate the degeneracy between the light of S1 and S2, which prevents successful lens models. We compute the overlap in the area as a percentage of the total source image area above a per-pixel SNR threshold of 2 and discard systems where at least one source has an overlap $>50\%$ of its total area. Using \textsc{Lenstronomy}, we computed conjugate points for each set of lensed images; these describe where the centre of a source galaxy maps to in each lensed image. This enabled us to quantify how many lensed images were present for each source and reject systems that have fewer than 2 conjugate points.

\section{Lens Modelling Methodology}\label{sec:modelling}

We used the open-source Bayesian modelling code \textsc{Herculens} \citep{galan_using_2022, galan_gordo_2024} to produce fully forward lens models from all of our mock images. \textsc{Herculens} was built on a JAX\footnote[4]{\url{https://github.com/jax-ml/jax}} framework, enabling it to efficiently run on GPUs with automatic differentiation of likelihood gradients. To produce multi-deflector plane models, we used extensions to the \textsc{Herculens} code as implemented in \cite{enzi_overconcentrated_2025}.

Due to the high dimensionality of the parameter space we want to explore, we must initialise our model before sampling the posterior distribution of model parameters. We optimise our model parameters through stochastic variational inference \citep[SVI, e.g.,][]{wingate_automated_2013, ranganath_black_2014}. With the chosen AdaBelief optimiser \citep{zhuang_adabelief_2020}, SVI maximises the evidence lower bound (ELBO), which is always less than or equal to the log of the Bayesian evidence given by a set of parameters. The ELBO is quantified as the expectation of the difference between the product of the prior and the likelihood and a multivariate Gaussian approximation to the true posterior, which can be computed efficiently. 

Our SVI optimisation is split into two stages: first, using a fully parametric model; second, using pixelated source surface-brightness reconstructions. In the first SVI stage, we penalise models that do not trace input conjugate points to the same point in the source plane and guide the optimiser to a focused solution. Starting with a simplified parametric model of the source light allows us to narrow the region of interest within the parameter space before introducing more complex pixelated reconstructions. Details of the parametric and non-parametric profiles used can be found in Sections \ref{sec:mass-profile} and \ref{sec:light-profile}. During both stages, we use 8 SVI chains, which perform 2000 iterations in the first stage and 10000 in the second stage. The median parameters from the last 1000 steps of each of the 4 best-fitting chains are used as starting points for sampling the posteriors.

We derive the posterior distributions of our lens model parameters through Markov chain Monte Carlo sampling. Our chosen Hamiltonian Monte Carlo (HMC) method uses an HMC-within-Gibbs hybrid approach with a No U-Turn Sampler \citep[NUTS;][]{homan_no-u-turn_2014}, where the NUTS serves as the inner kernel inside an outer Gibbs kernel \citep{krawczyk_ckrawczykmultihmcgibbs_2024}. Two blocks of parameters for the mass and light parameters are updated conditionally on one another in sequence, allowing their joint distributions to be approximated. We initialise 4 independent chains from our optimal SVI outputs and perform a warm-up of 2000 steps before sampling for a further 2000.

\subsection{Mass profiles}\label{sec:mass-profile}

The chosen mass profile for the lens is consistent across all modelling stages and is selected to be identical to that of the ground-truth SIE or EPL. We make this choice because we wish to examine the bias introduced by S1 specifically and assume a scenario where degeneracy in the mass model for the lens can be broken with additional information, such as velocity dispersion mapping. SIE and EPL parameterisations are often the standard choice for strong lensing analyses of elliptical galaxies \citep[e.g.,][]{bolton_sloan_2008, shajib_dark_2021, tan_project_2024, hogg_line--sight_2026}. This is consistent with the mass distribution of strong lenses, which are typically luminous red elliptical galaxies \citep[e.g.,][]{bolton_sloan_2008_VII, shu_sloan_2017, mukherjee_seagle_2018}.

Lensed source galaxies are not typically massive ellipticals and are instead star-forming galaxies with lower masses and more complex morphologies \citep[e.g.,][]{newton_sloan_2011, sonnenfeld_sl2s_2013, dye_modelling_2018}. However, less is known about the mass distributions of sources and simulated populations are often limited to smooth mass profiles \citep{collett_population_2015, korytov_cosmodc2_2019}. Therefore, we approximated the true mass distribution of S1 as an NFW + Sérsic and modelled it as an elliptical galaxy. We primarily explore models using a singular isothermal sphere (SIS) for the mass of S1. We also test extensions in the form of an SIE and EPL, which introduce ellipticity and a variable power-law slope, respectively. It should be noted that the difference between the model profile and a real source galaxy may be greater than what our ground truth allows.

The priors for each parameter we infer are listed in Table \ref{tab:priors}. Henceforth, we denote the Einstein radius of the lens as $\theta_{\rm E,l}$, and for an isothermal profile, this is identical to the definition of $\theta_{\rm E}^{\rm s1}$ (Section \ref{sec:criteria}). $\theta_{\rm E,s1}$ is the Einstein radius of a lens and source with redshifts of $z_{\rm s1}$ and $z_{\rm s2}$ for an isothermal profile. All Einstein radii are given in terms of arcseconds, as are the coordinates $x_0$ and $y_0$. Ellipticity is described in terms of dimensionless components $e_1$ and $e_2$, and residual shear is described by dimensionless components $\gamma_1$ and $\gamma_2$.

\setlength{\tabcolsep}{2pt}
\begin{table}
    \centering
    \begin{tabular}{|r|lccc|}
        \textbf{Component} & \textbf{Parameters} & \multicolumn{3}{c}{\textbf{Prior}} \\
        \hline
        SIS (S1) & $\theta_{\rm E}$ & $\mathcal{U}$[0, 0.8] & \\
        & $x_0$, $y_0$ & $\mathcal{N}$[Light centre, 0.01] & \\
        \hline
        SIE (S1) & $\theta_{\rm E}$ & $\mathcal{U}$[0, 0.8] & \\
        & $e_1$, $e_2$ & $t\mathcal{N}$[0, 0.25, -0.5, 0.5] & \\
        & $x_0$, $y_0$ & $\mathcal{N}$[Light centre, 0.01] & \\
        \hline
        EPL (S1) & $\theta_{\rm E}$ & $\mathcal{U}$[0, 0.8] & \\
        & $\gamma$ & $\mathcal{U}$[1.2, 2.8] & \\
        & $e_1$, $e_2$ & $t\mathcal{N}$[0, 0.25, -0.5, 0.5] & \\
        & $x_0$, $y_0$ & $\mathcal{N}$[Light centre, 0.01] & \\
        \hline
        SIE (lens) & $\theta_{\rm E}$ & $\mathcal{U}$[0, 4] & \\
        & $e_1$, $e_2$ & $t\mathcal{N}$[0, 0.25, -0.5, 0.5] & \\
        & $x_0$, $y_0$ & $t\mathcal{N}$[0, 0.1, -0.2, 0.2] & \\
        \hline
        EPL (lens) & $\theta_{\rm E}$ & $\mathcal{U}$[0, 4] & \\
        & $\gamma$ & $\mathcal{U}$[1.2, 2.8] & \\
        & $e_1$, $e_2$ & $t\mathcal{N}$[0, 0.25, -0.5, 0.5] & \\
        & $x_0$, $y_0$ & $t\mathcal{N}$[0, 0.1, -0.2, 0.2] & \\
        \hline
        Shear (lens) & $\gamma_1$, $\gamma_2$ & $\mathcal{U}$[-0.2, 0.2] & \\
        \hline
        Distance ratio & $\beta$ & $\mathcal{U}$[0.5, 1] & or & $\mathcal{U}[\frac{2}{9}, 1]$ \\
    \end{tabular}
    \caption{Mass-model parameters and their priors for each mass profile used. In addition, two priors on the distance ratio are listed, the second of which is for a subset of systems with more extreme distance ratios, with true values outside the standard prior range.}
    \label{tab:priors}
\end{table}

\subsection{Light profiles}\label{sec:light-profile}

In our simulated observations, we assumed Poisson-limited lens subtraction, omitting lens light but including the Poisson noise left after a lens subtraction. Therefore, within our lens models, we did not require a lens light model, but we accounted for the leftover noise using a lens light variance map, computed alongside the simulated observation, added to our model variance.

During our initial SVI optimisation, previously described, we used two source light models. We started by using multi-Gaussian expansion (MGE) to describe the surface brightness of each source. Our MGEs are composed of three elliptical Gaussian components, each with fixed standard deviations logarithmically spaced between $0.001\arcsec$ and $0.2\arcsec$. The amplitude, ellipticity and position of each MGE are fit during the SVI optimisation. In the second stage, we move to a non-parametric pixelated model initialised by our best-fit MGEs. Pixelated surface brightnesses are modelled through a Gaussian process with covariances governed by a Matérn power spectrum with three hyperparameters that control the amplitude, scale length and regularisation. Further details of this implementation can be found in other works using the same approach \citep[e.g.,][]{enzi_overconcentrated_2025, li_disentangling_2026}. Our initial values are determined from the best-fit power spectrum parameters to our optimised MGEs. The hyperparameters are then inferred simultaneously with our mass-model parameters.

\subsection{Model Inspection}\label{sec:inspec}

Instead of excessively fine-tuning our modelling set-up on a per-system basis, we performed a visual inspection to determine whether each model was a success or a failure. This ensures that the recovered values presented pertain to reasonable models that have at least two independent chains that have converged on the same solution and are not obviously unphysical. We also ensured that the source plane reconstructions of the two sources match those of the ground truth, confirming that the sources had been delensed correctly. Finally, we checked whether we achieved a reasonable level of residuals between the mock observations and model reconstructions, permitting some deviations within the source images, which may occur due to the lack of supersampling. Examples of models classed as success and failure can be found in Appendix \ref{sec:appB}.

\section{Results}\label{sec:results}

\subsection{Jackpot-like}

We present models of 110 Jackpot-like systems, for all of which we obtained posterior distributions with at least two converged HMC chains. Our mass models consisted of an EPL + shear for the lens and SIS for S1 using the priors listed in Table \ref{tab:priors}. We were able to achieve a model success rate near $100\%$ with only a few failures in edge cases. From the posteriors of each of our lens models, we computed summary statistics to enable large-scale model comparison. These statistics include the median and the 16th and 84th percentiles that bound the 68\% credible region for each model parameter; thus, any inferred values are quoted as such henceforth.

\begin{figure}
    \centering
    \includegraphics[width=1\linewidth]{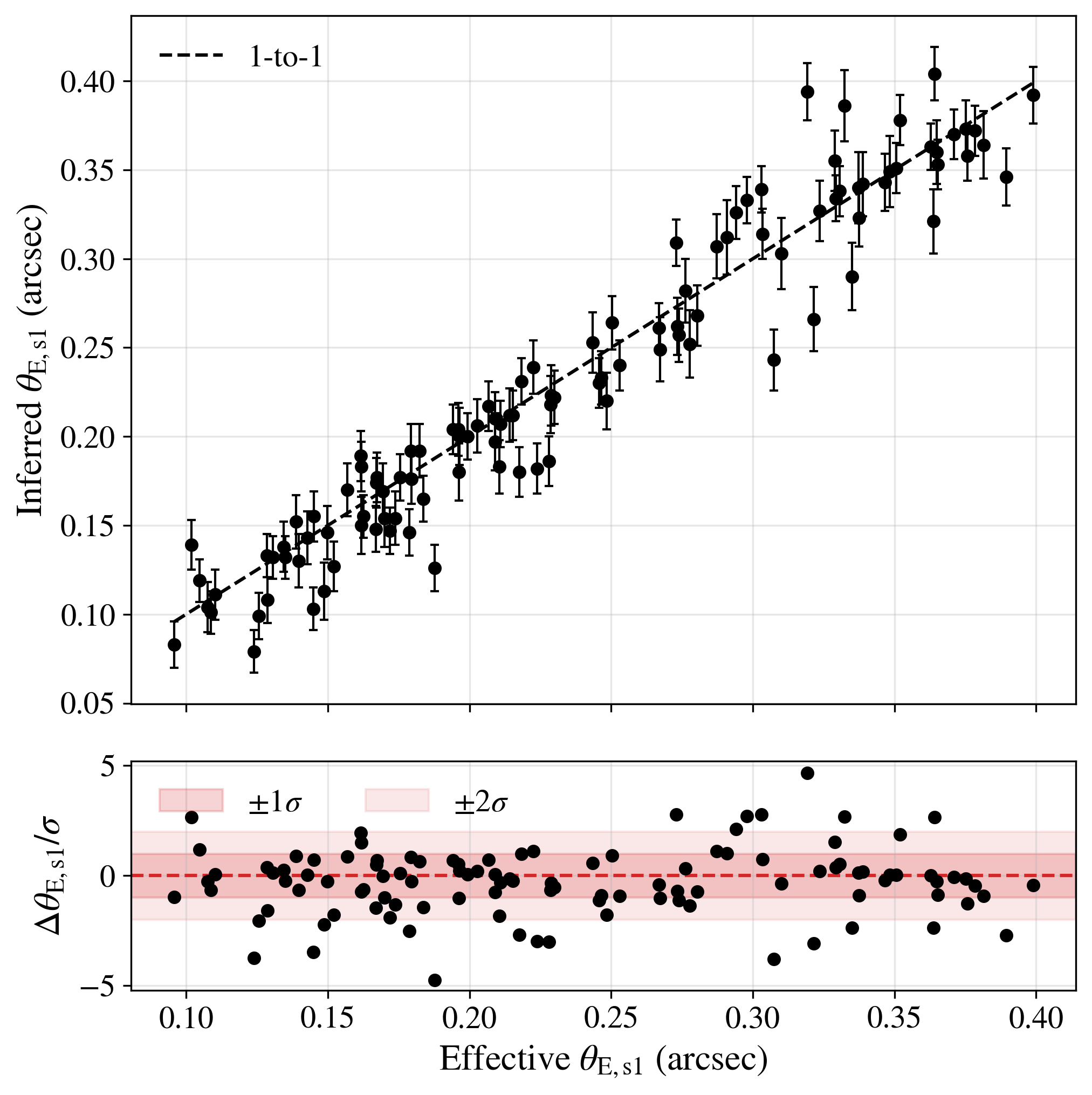}
    \caption{Top: Inferred Einstein radius $\theta_{\rm E,s1}$ of the SIS parameterisation of source 1 vs effective Einstein radius $\theta_{\rm E,s1}$ of source 1 computed from the ground-truth mass profile of the 110 Jackpot-like compound lenses modelled. Error bars represent the standard deviations of the posterior parameter distributions. The dashed 1-to-1 line represents where points would lie if they were in complete agreement. Bottom: Normalised residuals between the inferred and effective Einstein radius. Shaded regions represent a difference of one and two standard deviations, respectively.}
    \label{fig:theta_E_SIS_s1}
\end{figure}

We compared how well we recovered the parameters of the lens, which do not change between simulations, and find that they are consistently recovered within $2\sigma$ of the truth. We also compared how well we recovered the mass model properties of S1. However, due to the different parameterisations used by our model and the ground-truth mass profile, there is not an exact one-to-one mapping. Instead, we can compute an effective Einstein radius from our simulated profile and compare this with the parameterised Einstein radius within our lens model. From our inferred vs effective Einstein radius measurements in Figure \ref{fig:theta_E_SIS_s1}, we see good agreement between our models and simulations across the entire range of Einstein radii tested.

\begin{figure}
    \centering
    \includegraphics[width=1\linewidth]{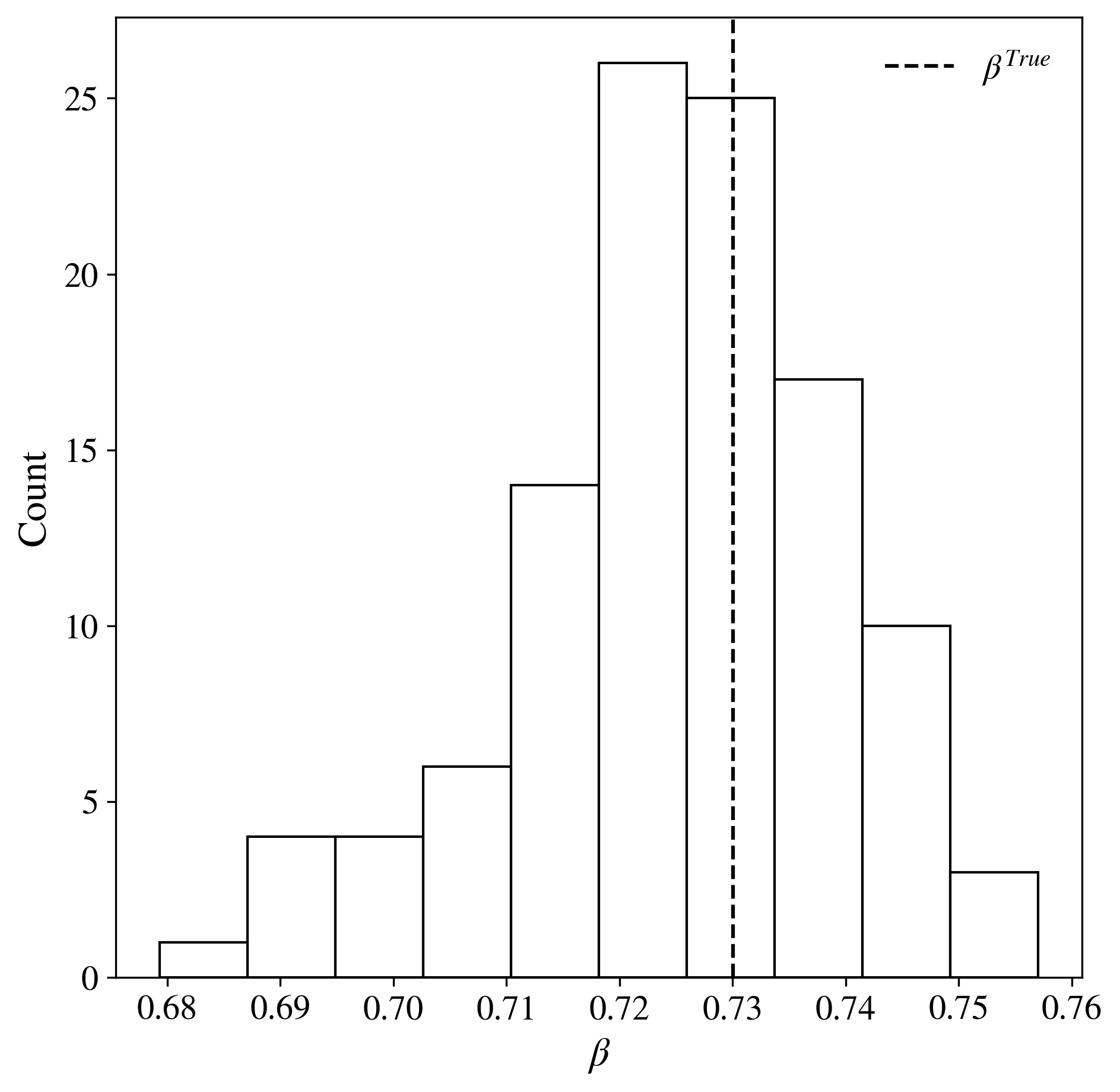}
    \caption{Distribution of inferred distance ratios $\beta$ of the 110 Jackpot-like compound lenses modelled. The dashed vertical line marks the true $\beta$ for every system.}
    \label{fig:beta_hist}
\end{figure}

The key cosmologically dependent parameter $\beta$ we inferred from our models shows good agreement with our constant ground-truth value. The distribution of the median values of $\beta$ from each model is shown in Figure \ref{fig:beta_hist}. Taking the combined median value from all models gives $\beta = 0.725 \pm 0.007$, which agrees with our ground-truth $\beta =0.730$ to within 1$\sigma$. However, as can be seen from the spread of values, there appears to be a bias towards smaller values of $\beta$. To investigate what could be driving the spread in inferred $\beta$, we performed Pearson tests for linear correlations between the normalised residuals of $\beta$ and the galaxy properties of S1, Table \ref{tab:correlation}.

The strongest and most significant correlation we found was between the normalised residuals of $\beta$ and $\log M_{*}/M_{\rm h}$ with a Pearson correlation coefficient of 0.675 and a p-value of $6\times10^{-16}$. $M_{*}/M_{\rm h}$ is derived from $M_*$ and $M_{\rm h}$ and thus correlated with them; therefore, it is likely the reason for the less significant correlations seen between them and $\beta$. In addition to the galaxy properties alone, there is a significant correlation between the normalised residuals of $\beta$ and the residual of $\theta_{\rm E,s1}$ with a Pearson coefficient of 0.357 and a p-value of $1\times10^{-4}$. The correlation reflects that the same bias is responsible for the correlation between $\beta$ and $M_{*}/M_{\rm h}$, as any change in $\beta$ will result in a change in $\theta_{\rm E,s1}$. 

\begin{table}
    \centering
    \begin{tabular}{rcc}
        \bf{Galaxy Property} & \bf{Correlation Coefficient} & \bf{p-value} \\
        \hline
        $\log M_*$ & 0.235 & 0.014 \\
        $\log M_{\rm h}$ & -0.510 & $1\times10^{-8}$ \\
        $\log M_*/M_{\rm h}$ & 0.675 & $6\times10^{-16}$ \\
        $\theta_{\rm E,s1}$ & 0.192 & 0.044 \\
        $q_*$ & -0.115 & 0.233 \\
        $q_h$ & 0.000 & 0.998 \\
    \end{tabular}
    \caption{Pearson correlation test results between the normalised residuals of $\beta$ inferred from our lens model featuring an SIS for source 1 and the galaxy properties of source 1: stellar mass $M_*$, halo mass $M_{\rm h}$, stellar-to-halo mass ratio $M_*/M_{\rm h}$, effective Einstein radius $\theta_{\rm E,s1}$, stellar axis ratio $q_*$ and halo axis ratio $q_h$. $M_*/M_{\rm h}$ and $\theta_{\rm E,s1}$ are properties derived from $M_*$ and $M_{\rm h}$. Included are Pearson correlation coefficients measuring the strength of correlation and p-values representing the probability that random uncorrelated data produces the same result.}
    \label{tab:correlation}
\end{table}

\begin{figure}
    \centering
    \includegraphics[width=1\linewidth]{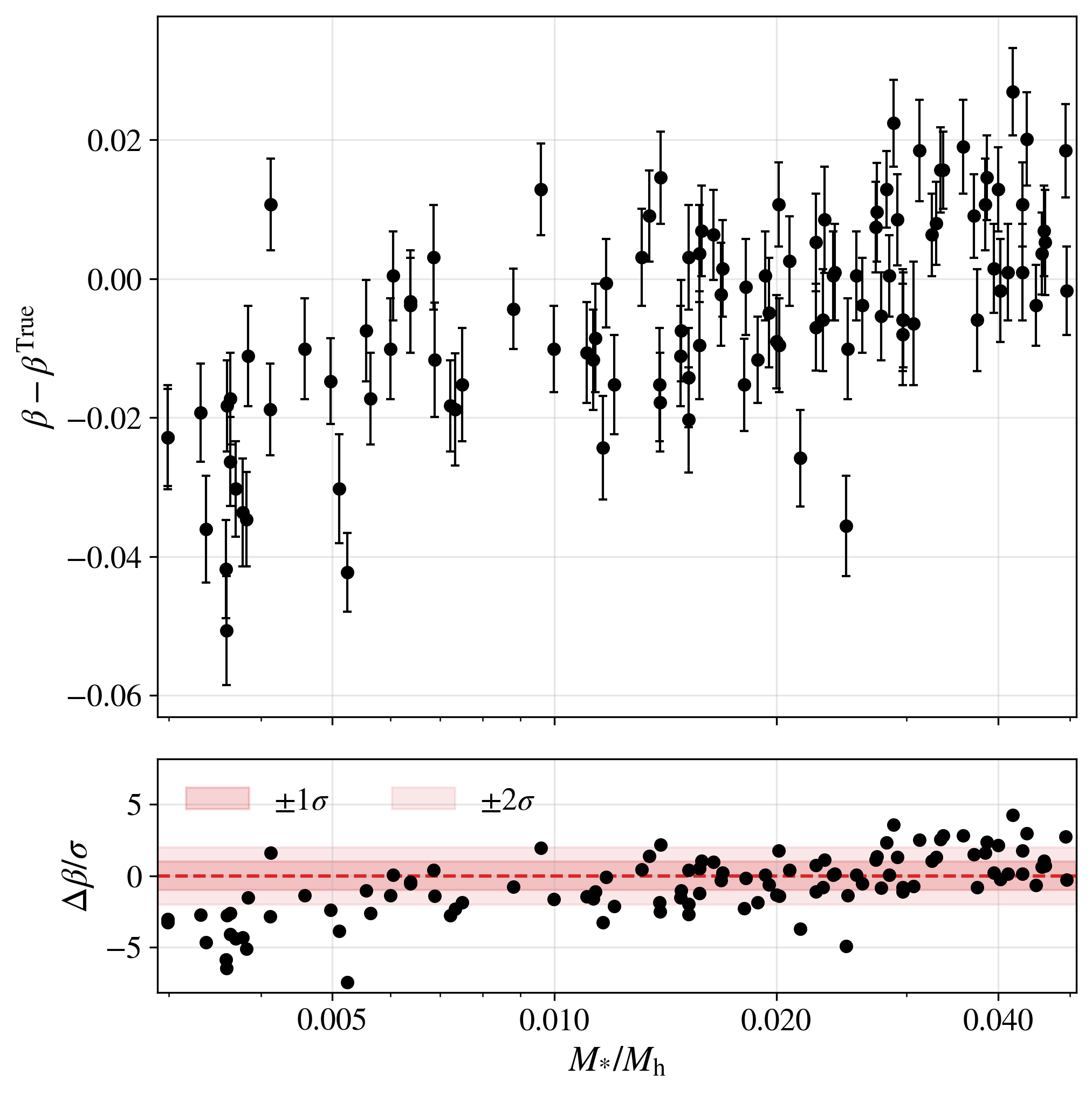}
    \caption{Top: Absolute residual between the inferred distance ratio $\beta$ and the true value $\beta=0.730$ vs stellar-to-halo mass of source 1 $M_{*}/M_{\rm h}$. Error bars represent the standard deviations of the posterior parameter distributions. Bottom: Normalised residuals between the inferred and true $\beta$. Shaded regions represent a difference of one and two standard deviations, respectively.}
    \label{fig:beta_MstarMh}
\end{figure}

The linear correlation between the normalised $\beta$ residuals and $\log M_{*}/M_{\rm h}$ is highlighted in Figure \ref{fig:beta_MstarMh}. For the smallest values tested, where $M_{*}/M_{\rm h}\simeq0.004$, the apparent offset between the recovered and true $\Delta\beta\simeq-0.03$ or $\simeq-4\sigma$. The reverse is true for larger values of $M_{*}/M_{\rm h}\simeq0.04$, but with less significance within the range of values tested. By splitting $M_{*}/M_{\rm h}$ into three regimes, we can identify the range over which $\beta$ is recovered accurately. There is a measurable negative bias for systems where $M_{*}/M_{\rm h} < 0.01$, for which the average recovered value is $\beta=0.713\pm0.007$. It follows that there is a positive bias when $M_{*}/M_{\rm h} > 0.03$, for which $\beta=0.738\pm0.007$. However, for systems where $M_{*}/M_{\rm h} \in [0.01,0.03]$, $\beta$ is more accurately recovered as $0.726\pm0.007$. This indicates that our mass model profile is a suitable approximation to the NFW + Sérsic ground truth for $M_{*}/M_{\rm h} \in [0.01,0.03]$ but less so outside this range.

\begin{figure*}
    \centering
    \includegraphics[width=1\linewidth]{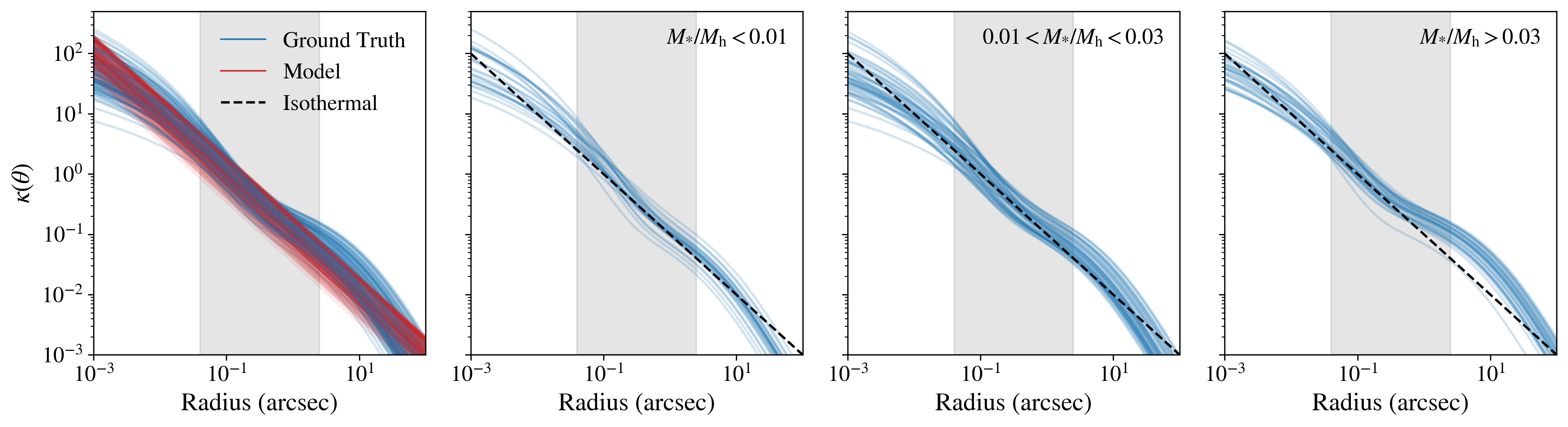}
    \caption{Azimuthally averaged radial convergence of source 1. Left: ground-truth and model-inferred convergence, in blue and red, respectively, of the 110 modelled jackpot-like systems. The ground truth follows an NFW + Sérsic profile and the model follows an SIS parameterisation. Middle left: ground-truth convergence for systems with $M_{*}/M_{\rm h} > 0.03$. Middle right: ground-truth convergence for systems with $M_{*}/M_{\rm h} \in [0.01,0.03]$. Right: ground-truth convergence for systems with $M_{*}/M_{\rm h} < 0.01$. Shaded regions represent the portion of the radial convergence profile our models are sensitive to across our simulated systems, bounded by the minimum and maximum separation between the images of S2 and the centre of S1 as they intersect the S1 plane. Dashed lines follow isothermal (SIS) profiles for reference.}
    \label{fig:fourpanel}
\end{figure*}

The left panel of Figure \ref{fig:fourpanel} compares the ground-truth to the model-inferred radial convergence $\kappa(\theta)$ of S1. Within the radii to which our lens models are most sensitive, there is a good agreement between the ground-truth and model profiles. In the four panels of Figure \ref{fig:fourpanel} we display the ground-truth profiles for systems with $M_{*}/M_{\rm h}$ split across three bins bounded by $M_{*}/M_{\rm h} = 0.01, 0.03$. These panels highlight the change in slope of the true profile as the stellar-to-halo mass ratio varies. We can see that the isothermal slope best approximates galaxies with $M_{*}/M_{\rm h} \in [0.01,0.03]$ in the middle right panel of Figure \ref{fig:fourpanel}, which is the most populated bin and is representative of typical galaxies.

In addition to the models presented, we produced models of the same simulated systems in which the SIS for S1 was replaced by an SIE. The SIE models produced posteriors that agreed with our SIS models, while producing a 1.3- and 1.6-times greater mean standard deviation in $\theta_{\rm E,s1}$ and $\beta$. The correlation between the normalised residuals of $\beta$ and $M_{*}/M_{\rm h}$ remained strongly correlated with a Pearson coefficient of 0.586 and a p-value of $2\times10^{-11}$. We also tested how an EPL for S1 influences the recovered model parameters on a subset of 45 simulated systems. The posteriors of the majority of the lens model parameters remained in agreement. However, significant bias was introduced into the recovered values of $\theta_{\rm E,s1}$ and $\beta$, which on average were overestimated by $2.99\sigma$ and underestimated by $3.45\sigma$, respectively. The increase in bias likely occurs because we have introduced too much flexibility into our model by allowing the density slope to vary, without having enough signal within our simulations to constrain this model. The priors used for the 

Our Jackpot simulations represent idealised compound lenses with two quadruply imaged sources with high signal-to-noise. From our models of these simulations, we identified the bias on $\beta$ as an internal MST of the first source introduced. There is a strong correlation between this bias and $M*/M_{\rm h}$. In the most extreme cases, where we push $M*/M_{\rm h}$ to 0.004, this results in as much as a 5$\sigma$ discrepancy between the true and inferred value of $\beta$. However, this collection of simulations is not representative of a varied population of compound lenses, like those being discovered by Euclid. Therefore, we present our results from models of a simulated sample of compound lenses detectable in the EWS in the following section.

\subsection{Euclid population}\label{sec:euclid_result}

From our \textsc{LenPop} catalogue, we found 165 detectable compound lenses ($\theta_{\rm E,s1}>0.1$) out of 2236 detectable DSPLs within $4200\ \rm{deg}^2$ of the sky ($30\%$ of the EWS) based on our criteria from Section \ref{sec:euclid-pop}. However, we discarded 15 of these which had $<2$ conjugate points, preventing them from being modelled. Therefore, we were left with 150 detectable systems suitable for modelling. Since our \textsc{LenPop} catalogue covers $4200 \ \rm{deg}^2$, we might expect roughly half the number of these systems within the aforementioned constraints to be detected by Euclid within DR1.

Of the 150 simulated Euclid DSPLs modelled, we obtained successful models for 74. We define what a successful model is in Section \ref{sec:inspec}. Unlike the Jackpot simulations, our model success rate depends on the SNR and morphology of image configurations, which vary greatly between simulations. This broader variation is the main reason we achieve successful models across $\sim50\%$ of our simulated Euclid lenses, compared to the Jackpot-like variations, which have a near-perfect success rate.

\begin{figure}
    \centering
    \includegraphics[width=1\linewidth]{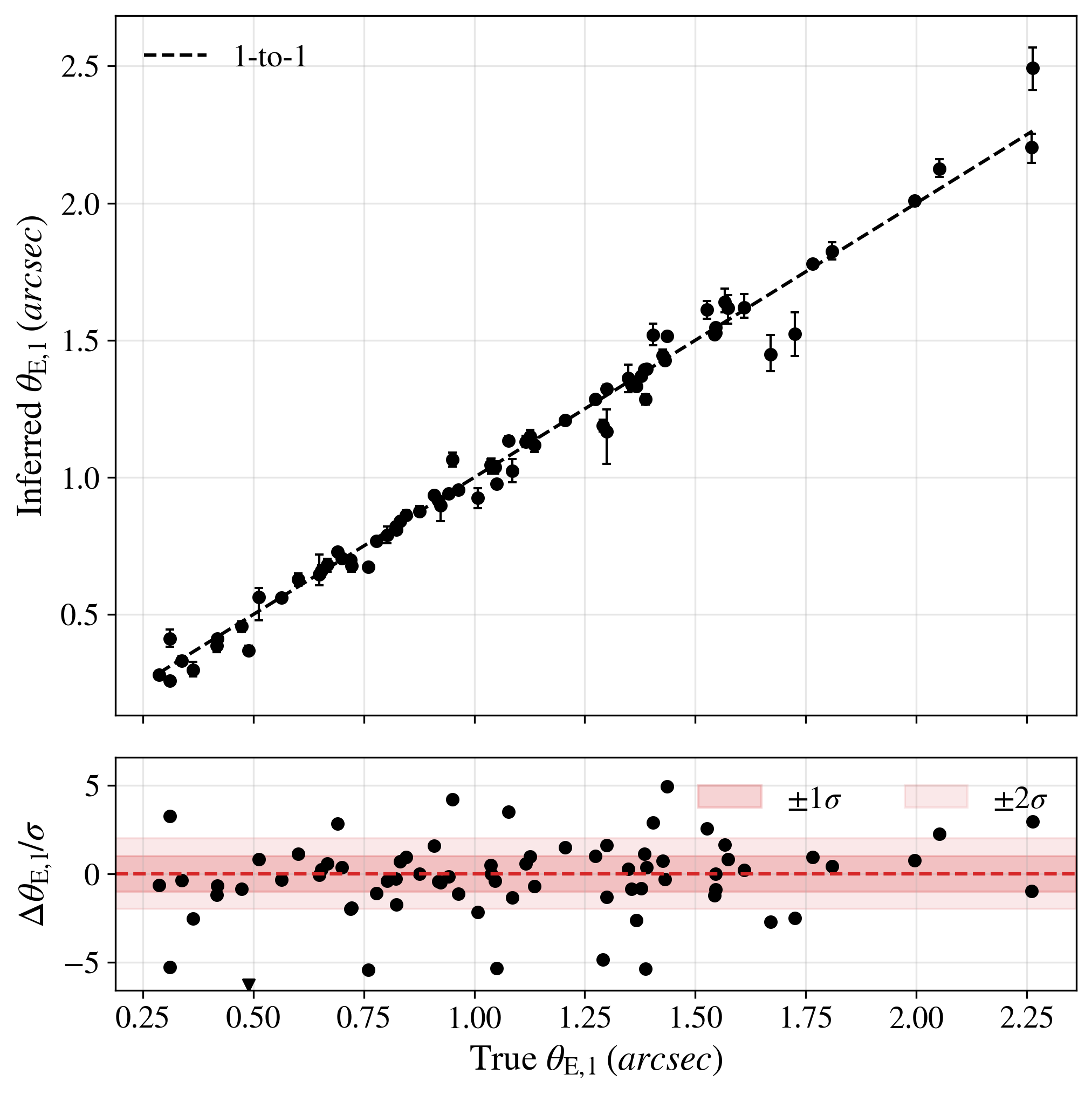}
    \caption{Top: Inferred Einstein radius $\theta_{\rm E,l}$ of the lens parameterised by an SIE vs true Einstein radius $\theta_{\rm E,l}$ of the SIE profile. Error bars represent the 16th and 84th percentiles of the posterior parameter distributions. The dashed 1-to-1 line represents where points would lie if they were in complete agreement. Bottom: Normalised residuals between the inferred and true Einstein radius. Shaded regions represent a difference of one and two standard deviations, respectively.}
    \label{fig:theta_E_l}
\end{figure}

We see a broad agreement between our inferred primary lens mass-model parameters and their true values. The Einstein radius of the lens $\theta_{\rm E,l}$ is consistently recovered across a range of values from 0.25\arcsec to 2.25\arcsec, shown in Figure \ref{fig:theta_E_l}. Only one model has a normalised residual much greater than 5$\sigma$. The strength of the inferred shear appears very high and could explain the large difference in $\theta_{\rm E,l}$. The precision to which $\theta_{\rm E,l}$ is inferred is approximately equal across the majority of systems, with few having larger uncertainties. 

\begin{figure}
    \centering
    \includegraphics[width=1\linewidth]{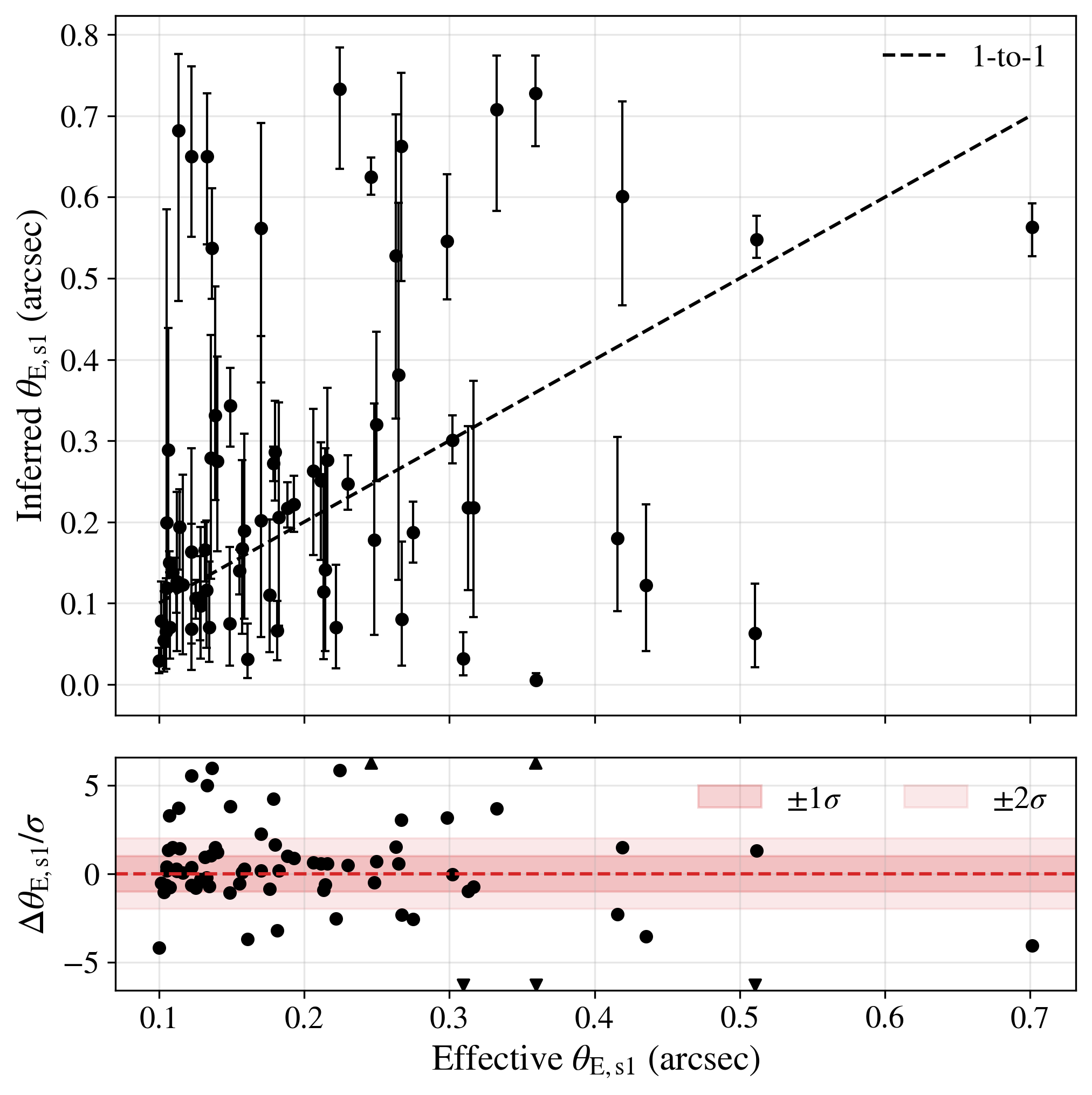}
    \caption{Top: Inferred Einstein radius $\theta_{\rm E,s1}$ of the SIS parameterisation of source 1 vs effective Einstein radius $\theta_{\rm E,s1}$ of source 1 computed from the ground-truth mass profile. Error bars represent the 16th and 84th percentiles of the posterior parameter distributions. The dashed 1-to-1 line represents where points would lie if they were in complete agreement. Bottom: Normalised residuals between the inferred and effective Einstein radius. Shaded regions represent a difference of one and two standard deviations, respectively.}
    \label{fig:theta_E_s1}
\end{figure}

We cannot consistently recover the effective Einstein radius of S1 $\theta_{\rm E,s1}$ with an uncertainty $<0.1$, as seen in Figure \ref{fig:theta_E_s1}. There appeared to be a weak negative correlation between the true $\theta_{\rm E,s1}$ and our normalised $\beta$ residuals, with a Pearson coefficient of -0.204 but a p-value of 0.08, indicating that it was not significant. However, due to large uncertainties in our inferred posteriors, $65\%$ of the recovered values were consistent within $2\sigma$ of the truth. Therefore, within the precision permitted by our models, $\theta_{\rm E,s1}$ was inferred without bias for the majority of our simulations. We did not find any correlations between the difference in our residuals of the inferred $\theta_{\rm E,s1}$ and any other model parameters or galaxy properties. This indicates that the information in our Euclid simulations was not sufficient to precisely constrain the properties of S1. Therefore, it is unlikely that we can identify whether any systematic bias is the result of S1.

\begin{figure}
    \centering
    \includegraphics[width=1\linewidth]{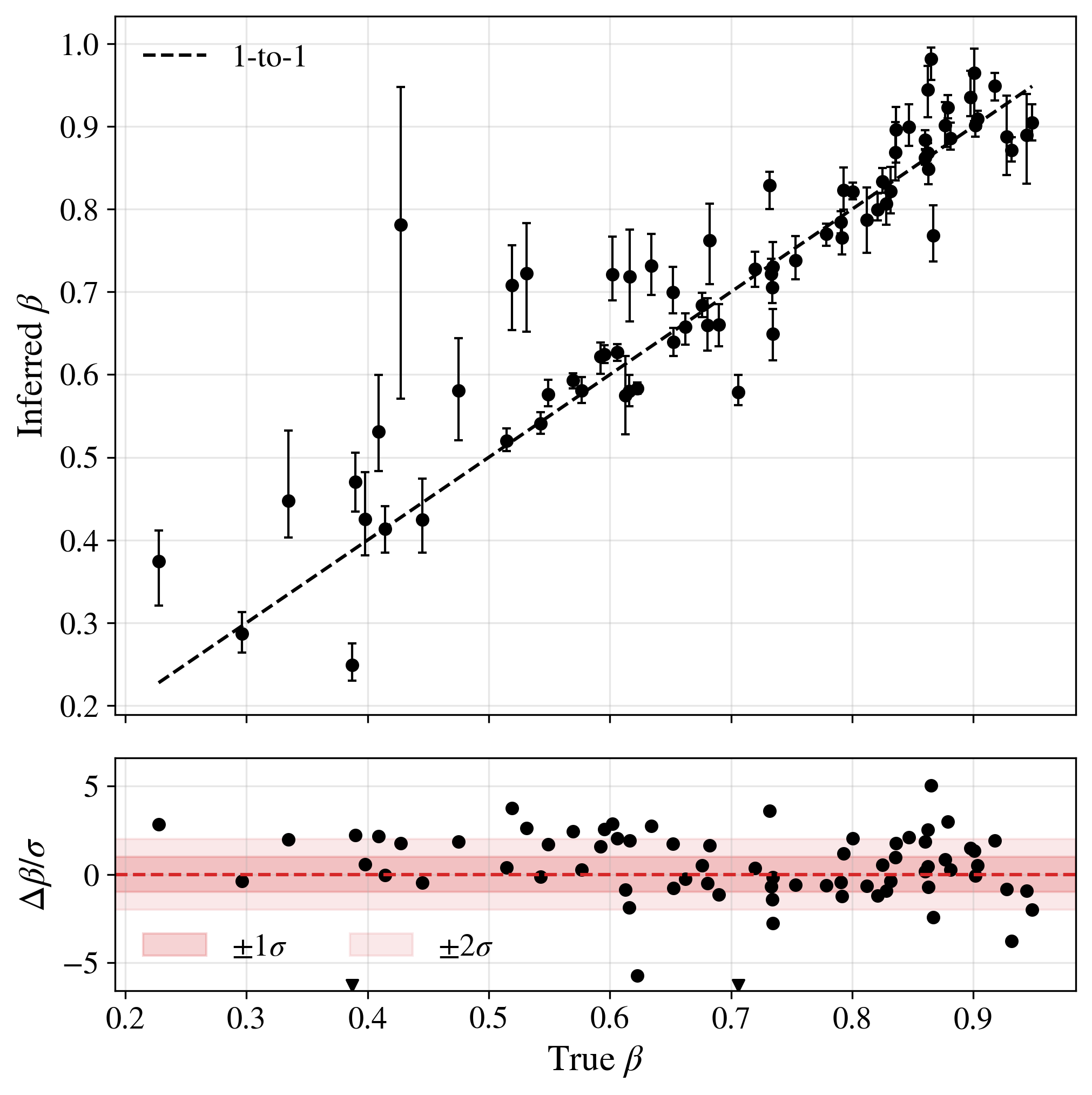}
    \caption{Top: Inferred distance ratio $\beta$ from each successful mock Euclid lens vs true $\beta$. Error bars represent the 16th and 84th percentiles of the posterior parameter distributions. The dashed 1-to-1 line represents where points would lie if they were in complete agreement. Bottom: Normalised residuals between the inferred and true $\beta$. Shaded regions represent a difference of one and two standard deviations, respectively.}
    \label{fig:beta}
\end{figure}

From our models, we were able to recover the distance ratio $\beta$ accurately in many cases but with significantly varying degrees of precision. In Figure \ref{fig:beta}, we see that there appears to be an increase in the $68\%$ credible interval as $\beta$ decreases. For simulations with true values of $\beta < 0.5$, this may in part be due to a prior volume effect. For the majority of systems, we used a $\beta \in [0.5, 1]$ uniform prior, as this provides enough freedom for the majority of models without allowing the model to become unphysical. However, for the extreme cases where $\beta$ falls below $0.5$ we use a larger uniform prior of $\beta \in [\frac{2}{9}, 1]$. The extreme cases are also most likely to be produced in systems where the separation between the images of S1 and the lens is nearing the full width half maximum (FWHM) of the PSF, resulting in a larger uncertainty for many of these systems. The model with the largest uncertainty on $\beta$ is an example where $\theta_{\rm E,l}$ is small ($\sim0.3$), and there is overlap and blending between the images of S1 and S2, leading to a poorly constrained model.

Using a Pearson correlation test, we do not identify any significant correlation between the normalised residuals of $\beta$ and any of the galaxy properties of S1, including $M_{*}/M_{\rm h}$. However, we do see a weak positive correlation (0.444) between the normalised residuals of $\beta$ and $\theta_{\rm E,s1}$ with a p-value of $1\times10^{-4}$. The cause of this correlation is likely an internal MST in which a decrease in deflection by S1 results in the angular distance of S2 increasing and $\beta$ decreasing.

Four of our 74 successful models recover a value of $\beta$ > 5$\sigma$ away from the truth, the two most significant of which have large effective $\theta_{\rm E,s1}$ comparable to $\theta_{\rm E,l}$. The models of these systems significantly underestimate $\beta$ and $\theta_{\rm E,s1}$. This suggests a strong degeneracy between the two parameters within these systems and aligns with the correlation found between the normalised residuals of $\beta$ and $\theta_{\rm E,s1}$. The remaining two have a significant overlap between the images of S1 and S2, allowing some of the light from one source to compensate for the other, which could explain the discrepancy in $\beta$. 

There are additional clusters of models within Figure \ref{fig:beta} that are offset from the 1-to-1 line. For a significant fraction of these systems that are biased upward and have larger uncertainty, one source often does not have a detectable counter image, either because the pair of detectable images are separated in angle by $< 120\deg$, or there is a complete overlap between the counter of one source and the images of the other source. These systems are highlighted in Figure \ref{fig:no_counter} of Appendix \ref{sec:appA}.

\subsubsection{The Gold Sample}\label{sec:gold}

In a real-world scenario, we might want to perform follow-up observations on a subsample of the most valuable DSPLs for cosmography, i.e. the gold sample. By obtaining higher-resolution (0.05\arcsec) imaging with a telescope such as HST, this would allow us to model those systems more precisely and tighten constraints on model parameters of importance such as $\beta$. We emulated this procedure by producing simulated HST observations of the gold sample and tested how well we could recover our model parameters.

To obtain our gold sample of DSPLs, we down-selected our successfully modelled systems using more restrictive constraints. We required that $\beta < 0.9$, $\theta_{\rm E}^{\rm s1} > 0.5$, either S1 or S2 had three detectable images, the overlap in area between the two sources was < $20\%$ above a pixel SNR threshold of 2, and we removed any systems where there is no detectable counter image or if it completely overlaps with another image, where we define an image pair to have a counter when it is subtended by an angle $>120^\circ$. These are in addition to all of our existing constraints listed in Section \ref{sec:criteria}. This left us with 20 compound lenses in our gold sample.

Unlike our previous model inspection methodology, described in Section \ref{sec:inspec}, used for the range of Euclid simulations, we take a different approach to the fixed model setup from Section \ref{sec:modelling} for our gold HST simulations. Because of the nature of the gold sample being small, we do not want to lose systems due to model failures. Therefore, we adjust our model setup on a per-system basis where required to produce a successful model, mimicking the steps taken to analyse a smaller lens sample. This included increasing the number of samples taken by the NUTS, adjusting conjugate points and replacing S1's SIS with an SIE.

\begin{figure}
    \centering
    \includegraphics[width=1\linewidth]{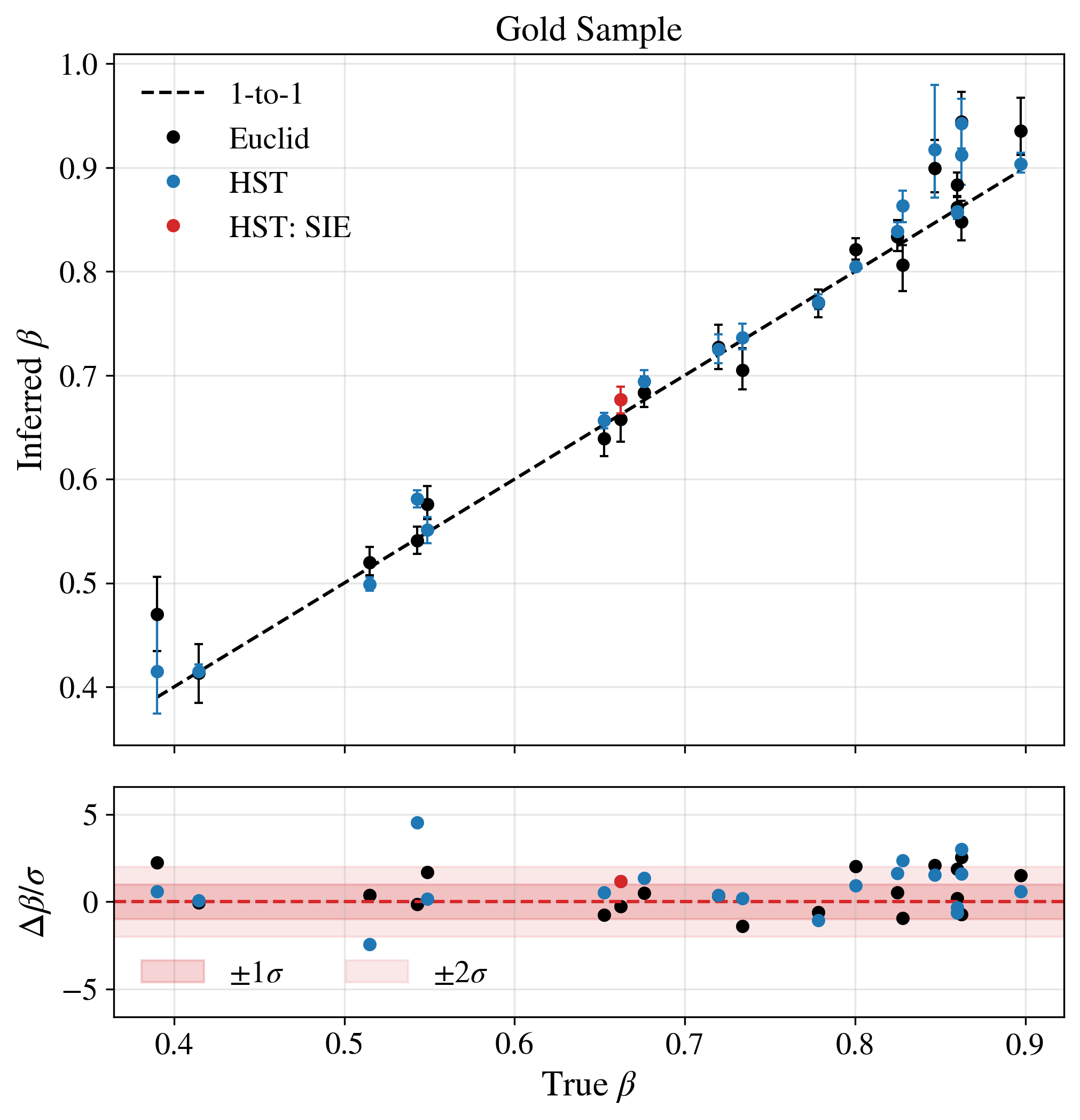}
    \caption{Top: Inferred vs true distance ratio $\beta$ from our gold sample of compound lenses. Black points represent the models of Euclid simulations and blue points represent the models of simulated HST observations. Error bars represent the 16th and 84th percentiles of the posterior parameter distributions. The dashed 1-to-1 line represents where points would lie if they were in complete agreement. Bottom: Normalised residuals between the inferred and true $\beta$. Shaded regions represent a difference of one and two standard deviations, respectively.}
    \label{fig:gold}
\end{figure}

We compare the inferred $\beta$ values from our Euclid models with the simulated HST observations of our gold sample in Figure \ref{fig:gold}. We do not detect any significant bias in the HST models over Euclid and see that the normalised residuals for the gold sample do not exceed 3$\sigma$ for the Euclid or HST models. The mean 1$\sigma$ credibility interval for the Euclid models was $0.021$, which reduced to $0.014$ when modelling the HST observations. Therefore, we are able to recover $\beta$ with comparable accuracy between the simulations but with greater precision from the HST simulations.

\subsubsection{Zigzag Lenses}

Einstein zigzags are a rare kind of compound lens in which additional images of sources beyond the first can form due to extreme compound lensing. For a DSPL, this occurs when the ratio of the deflection by S1 over the lens is greater than $1-\beta$ but less than 1, resulting in higher multiplicity and allowing up to 16 lensed images of S2 \citep{collett_compound_2016}. The zigzag configuration is more sensitive to the mass distribution of S1 because S2 lies within its caustics. Therefore, zigzags could provide tighter model constraints on S1 but might also demand more model complexity.

\begin{figure}
    \centering
    \includegraphics[width=1\linewidth]{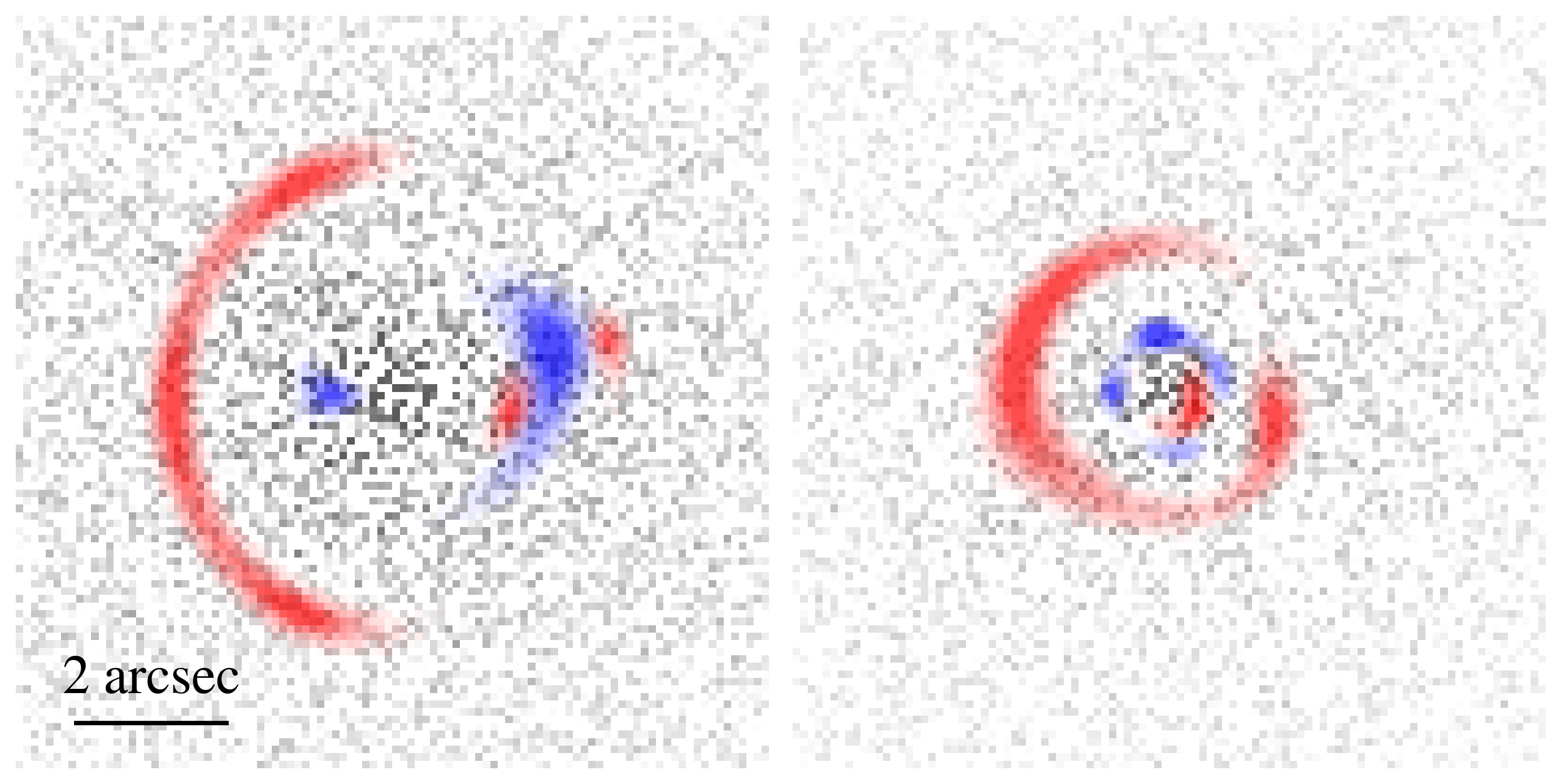}
    \caption{Simulated false-colour Euclid imaging of the two Einstein zigzags within our gold compound lens sample. The light from source 1 is highlighted in blue, and the light from source 2 is highlighted in red. Lens light is omitted, but the residual Poisson noise from the lens remains, in addition to background noise, which are displayed in greyscale.}
    \label{fig:zigzag}
\end{figure}

We identified a total of 5 zigzag lenses in our 150 detectable simulated systems. Of these, we obtained successful Euclid models for two, which are included in our gold sample. One of these has 5 images corresponding to a single source (left panel of Figure \ref{fig:zigzag}). Modelling the mock HST observations required the use of an SIE profile for S1 to trace the conjugate points back to a single origin in the source plane. The second zigzag consisted of a pair of images and a third image of S2 within the Einstein cross produced by S1 (right panel of Figure \ref{fig:zigzag}). Here, an SIE was not required to produce a plausible model. However, this produced the most biased value of $\beta$ from our gold HST sample. Consequently, we remodelled this simulation using an SIE for S1, but instead of reducing the bias in $\beta$, it greatly increased it, as seen in Figure \ref{fig:gold}. The reason why this simulation in particular is biased is unknown, but could be due to a strong degeneracy between $\theta_{\rm E,s1}$, which is over-predicted, and $\beta$. For our other zigzag, $\theta_{\rm E,s1}$ is more significantly over-predicted, yet the resulting $\beta$ is in agreement with the truth. This difference is likely due to the image configuration. When S1 is closer to the centre of the lens, it is harder to distinguish between an increase in $\theta_{\rm E,s1}$ and a decrease in $\beta$. Whereas a doubly imaged S1 produces a less radially symmetric deflection that is not degenerate with $\beta$.

Einstein zigzags are complex systems for which our lens modelling procedure was not optimised to model. There were 3 detectable zigzags within our Euclid sample that did not produce successful models, leaving us with a sample of 2. With a limited simulated sample, we are unable to determine the benefits or challenges they may present. However, answering these questions is beyond the scope of this paper and is left for future work.

\subsection{Cosmological Constraints}

By combining our posterior distance ratios with the addition of redshifts $z_{\rm l}$, $z_{\rm s1}$ and $z_{\rm s2}$ for each simulated compound lens, we inferred constraints on the dark energy equation of state parameter $w_0$ and the matter density parameter $\Omega_{\rm m}$, for a flat $w$CDM cosmology. Our inferred lens model parameter $\beta$ is related to $w_0$ and $\Omega_{\rm m}$ through Equations \ref{eq:beta}-\ref{eq:E(z)}.

\begin{figure}
    \centering
    \includegraphics[width=1\linewidth]{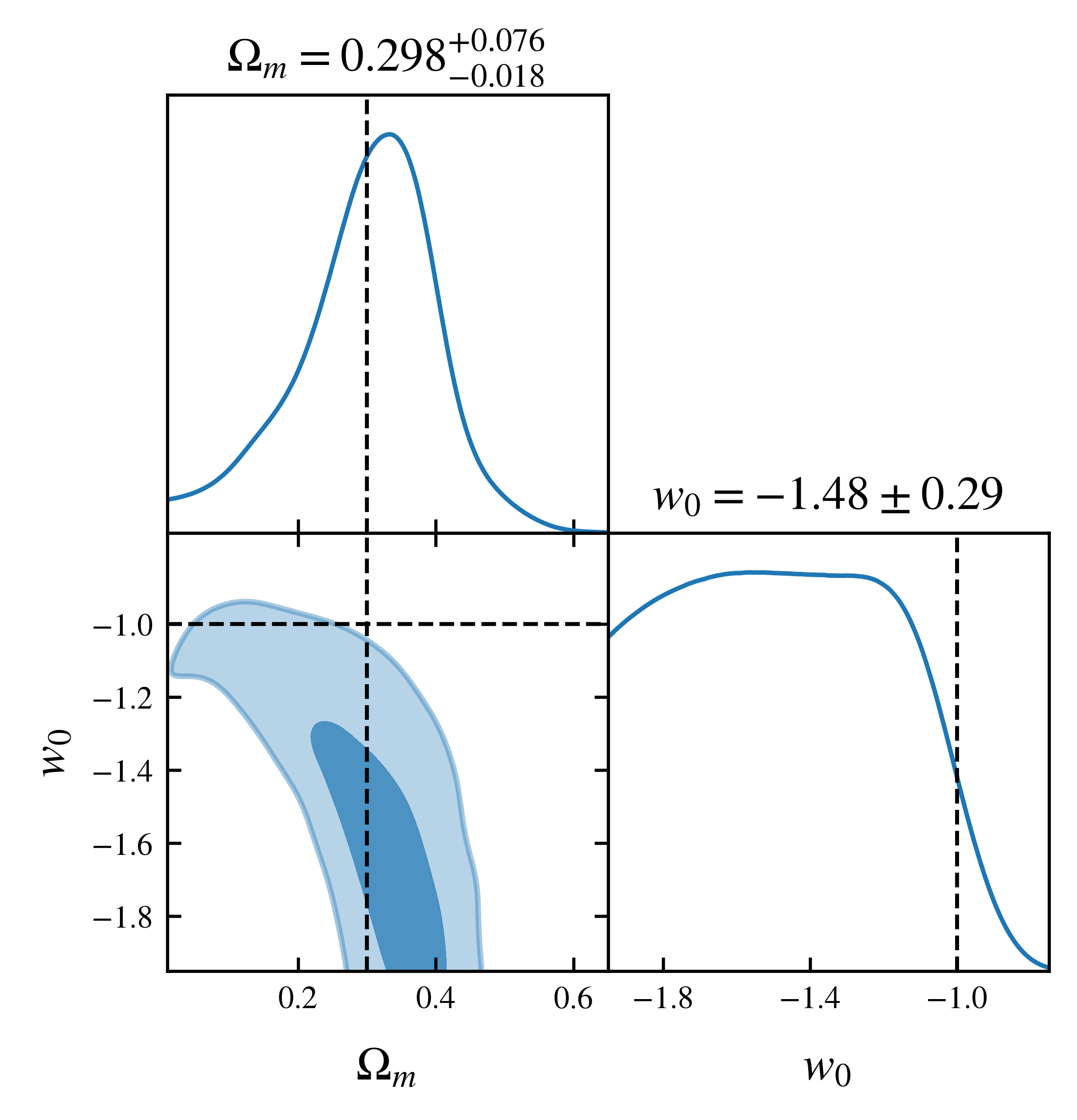}
    \caption{2D distributions of $w_0$ and $\Omega_{\rm m}$ under the assumption of a flat $w$CDM cosmology derived from the combined posteriors of 19 of our gold compound lenses (after removal of one system) with simulated Euclid observations. Shaded contours represent the $1\sigma$ and $2\sigma$ credible regions, and histograms represent the 1D distributions of $w_0$ and $\Omega_{\rm m}$, for which median values are quoted with $1\sigma$ uncertainties. The ground-truth parameter values ($w_0=-1, \Omega_{\rm m} = 0.3$) are highlighted by the dashed lines.}
    \label{fig:euclid_gold_wcdm}
\end{figure}

\begin{figure}
    \centering
    \includegraphics[width=1\linewidth]{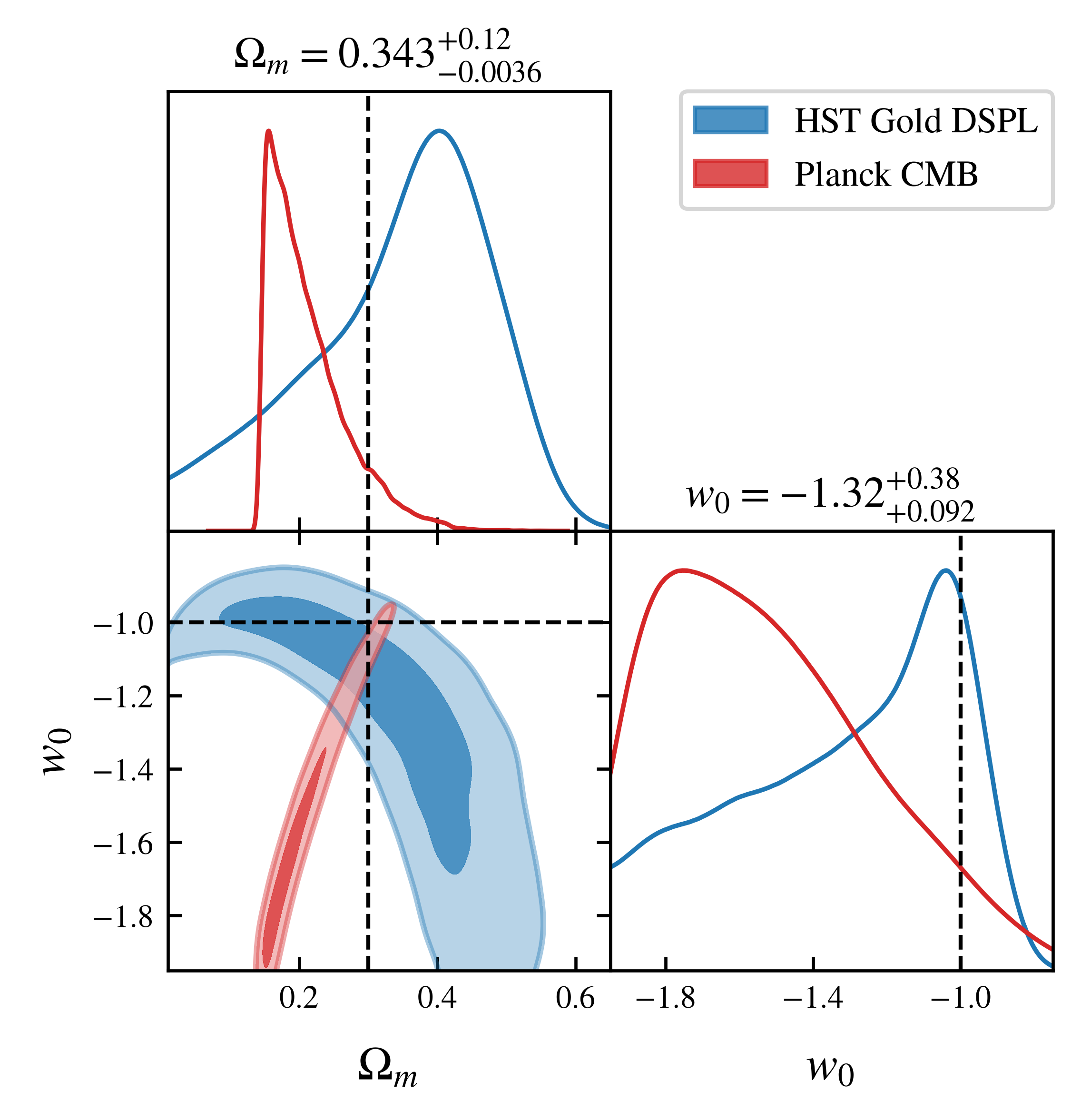}
    \caption{2D distributions of $w_0$ and $\Omega_{\rm m}$ under the assumption of a flat $w$CDM cosmology derived from the combined posteriors of 19 of our gold compound lenses (after removal of one system) with simulated HST observations (blue) and CMB measurements from \citet{planck_collaboration_planck_2020} (red). Shaded contours represent the $1\sigma$ and $2\sigma$ credible regions, and histograms represent the 1D distributions of $w_0$ and $\Omega_{\rm m}$, for which median values inferred from the gold lenses are quoted with $1\sigma$ uncertainties. The ground-truth parameter values ($w_0=-1, \Omega_{\rm m} = 0.3$) of the simulated systems are highlighted by the dashed lines.}
    \label{fig:hst_gold_wcdm}
\end{figure}

We produced constraints on $w_0$ and $\Omega_{\rm m}$ from the posterior distributions of $\beta$ acquired by our lens models. We sampled the posterior within the $w_0-\Omega_{\rm m}$ parameter space using the HMC NUTS. We used 4 independent chains, each performed a warm-up of 1000 iterations followed by 10000 samples. Uniform priors were placed on $w_0$ and $\Omega_{\rm m}$ with the boundaries [-2.0, 0.0] and [0.0, 1.0], respectively. We approximated the posterior distributions of $\beta$ using a kernel density estimate (KDE), allowing for non-Gaussianities, from which our cosmological constraints were sampled. The 2D $1\sigma$ and $2\sigma$ credible regions along with the 1D distributions of $w_0$ and $\Omega_{\rm m}$ are displayed in Figures \ref{fig:euclid_gold_wcdm} and \ref{fig:hst_gold_wcdm}.

When taking the posterior $\beta$ from all our successfully modelled Euclid simulations, we did not obtain physically plausible constraints on $w_0$ and $\Omega_{\rm m}$. This was due to systems with a biased value of $\beta$ pulling the credible region to extreme values of the $w_0-\Omega_{\rm m}$ parameter space. Taking only the posterior from our gold sample, we were able to recover a cosmology for which there is a tension of $\sim2\sigma$ with the ground truth (Figure \ref{fig:euclid_gold_wcdm}). Using simulated HST observations of this gold sample, we improved the precision of our $\beta$ posteriors, reducing the mean standard deviation by $37\%$, resulting in more accurate cosmological constraints for which the ground truth is $\sim 1\sigma$ away and reduced our $1\sigma$ uncertainties on $w_0$ by $50\%$ (Figure \ref{fig:hst_gold_wcdm}). Here, our gold sample contains 19 systems, rather than the 20 from Section \ref{sec:gold}. We removed our most biased HST model posterior, which occupies an unphysical region of the $w_0-\Omega_{\rm m}$ parameter space, much greater than $3\sigma$ from our ground truth $w_0 = -1$ and $\Omega_{\rm m} = 0.3$. In Figure \ref{fig:hst_gold_wcdm}, we show the constraints on $w$CDM provided by CMB measurements \citep{planck_collaboration_planck_2020} for reference and to highlight the orthogonality between DSPLs and the CMB.

\section{Discussion}\label{sec:discussion}

Across our Jackpot simulations, we find that we can recover the cosmologically informative distance ratio $\beta$ with a systematic bias of no more than $7\%$ for an individual. The source of this bias is dominated by compound lensing and the influence of the mass of S1. Our maximal bias occurred when replacing the first source of the Jackpot with galaxies that have a stellar-to-halo mass $M_{*}/M_{\rm h}\simeq0.005$. Through the difference between the true mass profile and a given model parameterisation, we have performed an effective test of how a physically motivated internal MST transform in the first source plane biases $\beta$. A transformation of this type had not been tested before this work, unlike more generalised MSTs \citep{schneider_can_2014}.

We examined two regimes based on data quality and inherent constraining power. The first of these used simulated HST-like observations of mock compound lenses based on the real Jackpot DSPL. This encapsulates the ideal scenario in which we have high-resolution follow-up observations of a DSPL that is well suited for cosmography because of its lensing configuration. Second, we attempted to quantify how any bias propagates to a larger, more varied population of DSPLs without dedicated follow-up observations, i.e. those that Euclid is now discovering. 

From the models of our HST simulations, we identified a bias that is strongly correlated with the $M_{*}/M_{\rm h}$ of S1 (Figure \ref{fig:beta_MstarMh}). It follows that this bias is caused by the internal MST between the ground-truth NFW + Sérsic and the selected SIS mass profiles. Varying $M_{*}/M_{\rm h}$ alters the average mass density slope due to the change in the relative contributions between the NFW and Sérsic profiles (Figure \ref{fig:fourpanel}). Due to the limitations of the SIS profile, the model cannot account for this change in the density slope. Therefore, the mismatch results in an effective MST that propagates to the inferred value of $\beta$. However, for values of $M_{*}/M_{\rm h}\in[0.01,0.03]$ we found that the bias is negligible compared to statistical model variance. 

In our models of the mock Euclid population, we do not detect a bias in $\beta$ that correlates with $M_{*}/M_{\rm h}$. This does not indicate that the same bias does not exist, but our mock Euclid observations do not contain sufficient information to identify this bias over other modelling uncertainties. This is highlighted by the lack of consistent recovery of $\theta_{\rm E,s1}$ in Figure \ref{fig:theta_E_s1}, which itself has a larger impact on the lensing observables, and implies that we cannot disentangle the influence of S1 from the uncertainty in our lens models. Although the measurement precision of $\beta$ improves with our models of the HST simulations of our gold sample of Euclid mocks, the correlated bias we found in the Jackpots is not detectable here, either due to inferior and variable SNR or the sample size being significantly smaller.

Strong lensing studies of early-type deflectors at $z\sim0.01-0.7$ have shown that the mean density slope $\gamma$ is consistent with that of an isothermal ($\gamma=2)$ profile with some intrinsic scatter $\sigma \simeq 0.15$ using lensing + stellar kinematic and lensing-only information \citep[e.g.,][]{auger_sloan_2010, lyskova_mass_2018, ritondale_resolving_2019, shajib_dark_2021, knabel_spatially_2026}. This is supported by spatially resolved stellar kinematic measurements of early-type galaxies \citep[e.g.,][]{cappellari_small_2015, poci_systematic_2017}, although with slightly steeper slopes. Since $\gamma$ is strongly correlated with $M_{*}/M_{\rm h}$, this explains why our bias in $\beta$ is negligible for $M_{*}/M_{\rm h}\in[0.01,0.03]$, where the NFW + Sérsic is well approximated by an SIS. 

Several studies using lensing and/or kinematics have investigated whether $\gamma$ evolves with redshift, many of which are compared by \cite{sahu_agel_2024}. However, there is no consensus on whether evolution exists up to $z\sim1$. A recent study using resolved stellar kinematics finds a decrease in slope consistent with lensing + kinematics studies, where the steeper slope becomes closer to isothermal as redshift increases \citep{mozumdar_magnus_2026}. For DSPLs, we must also consider that S1 is not necessarily an early-type galaxy, unlike the majority of strong lenses and therefore may not abide by the isothermal approximation.

Here we have examined only the systematic bias an internal MST within S1. However, MSTs not considered in this work, such as the internal MST within the lens \citep[e.g.,][]{collaboration_tdcosmo_2026} and external MST due to LOS perturbers \citep{johnson_line--sight_2025} can introduce additional modelling uncertainty but can be mitigated with stellar kinematic information and external convergence measurements. LOS perturbers have been shown to add $1.5\%$ measurement uncertainty to $\beta$, even when utilising galaxy number counts to constrain the external mass sheets \citep{johnson_line--sight_2025}. These effects can be more easily constrained with independent observations, including stellar kinematic measurements \citep[e.g.,][]{sheu_tdcosmo_2026} and galaxy number counts \citep[e.g.,][]{johnson_tdcosmo_2026}. Consequently, the internal MST within S1 remains a key source of error that must be accounted for. More complex mass profiles can also introduce additional degeneracy, such as the EPL with a free power-law density slope, which can partially imitate an MST (Johnson et al. in prep.), widening the credible region of $\beta$. All such sources of error must be quantified and propagated if DSPLs are to become a robust probe of cosmology.

From existing surveys, we have only $\mathcal{O}$(10) known DSPLs, but the imminent release of Euclid DR1 will bring the number of candidates close to 100 \citep[][Euclid Collaboration in prep.]{li_euclid_2025}. However, from our findings, the usable (gold) sample of systems is expected to be an order of magnitude smaller. Consequently, we might expect that constraints comparable to those presented in Figures \ref{fig:euclid_gold_wcdm} and \ref{fig:hst_gold_wcdm}, plus the additional uncertainties mentioned in the previous paragraph, are achievable with the DSPLs discovered in Euclid DR1 after spectroscopic follow-up. If we consider that the DR1 lenses were identified by visual inspection without lens subtraction (Euclid Collaboration in prep.), it is likely that a large fraction of the $\sim 100$ DR1 DSPLs will fit the criteria for our gold sample. 

\section{Conclusions}\label{sec:conclusion}

DSPLs are promising potential probes of dark energy that can leverage distance ratios to constrain cosmological parameters. However, one of the most prominent degeneracies in strong lensing, caused by the mass-sheet transformation, is not only persistent in DSPLs but is further complicated when we consider the presence of a second deflector plane. Compound lenses consist of two or more deflectors within independent redshift planes, allowing for MSTs in each. The influence of an internal MST within the intermediate deflector (S1) plane has scarcely been considered before, yet has significant implications for how DSPLs should be treated.

In this paper, we have investigated a bias unique to compound lenses through complete lens model reconstructions of simulated compound lenses within a Bayesian framework. We examined how an internal MST of S1 altered the inferred distance ratio parameter $\beta$ across two regimes. In the first regime, we explored simulations based on the real DSPL, the ``Jackpot'' J0946+1006 \citep{gavazzi_sloan_2008, collett_cosmological_2014}. In the second, we produced a more varied population of simulated Euclid compound lenses. In both sets of simulations, we focused on varying the mass distribution of S1 using an NFW + Sérsic parameterisation, allowing for a more complex galaxy profile than is typically used in lens modelling \citep[e.g.,][]{bolton_sloan_2008, hogg_line--sight_2026}. Performing lens model inference using \textsc{Herculens}, we produce optimised lens model reconstructions while sampling the posterior distributions of lens model parameters. In our models, we approximated S1 using isothermal mass profiles and examined any resulting bias in the cosmologically-dependent distance ratio parameter $\beta$.

From our simulated variations of the Jackpot, we introduced a maximum bias of $\sim7\%$ to the inferred value of $\beta$ due to the difference between the true and model-parameterised mass distribution of S1. This bias is strongly correlated with $M_*/M_{\rm h}$ of S1 and only extreme values of $M_*/M_{\rm h}$ produced a bias $> 3\sigma$. Therefore, through compound lensing, DSPLs are sensitive to changes in the mass profile of S1 that directly influence the value of $\beta$. Per cent-level deviations in $\beta$ can propagate to much larger biases in cosmological parameters \citep{schneider_can_2014}. Consequently, S1 must be treated with care if we wish to use DSPLs as a future probe of dark energy.

In our simulated sample of compound lenses detectable by Euclid, we were unable to detect the same bias in the recovered values of $\beta$. Due to limitations in the sensitivity of our simulated observations to the deflection caused by S1, we were unable to recover $\theta_{\rm E,s1}$ consistently, thus preventing us from constraining the effect of compound lensing. In turn, we did not identify any correlated bias in $\beta$ with the properties of S1. However, the scatter in recovered $\beta$ remained greater than that produced by statistical uncertainty. We selected a gold subset of our Euclid simulations based on criteria that reduce the spread in inferred values of $\beta$. By selecting systems with larger distance ratios and Einstein radii, greater image multiplicity, and image separation, we can improve the accuracy of our models and remove biased systems, but at the cost of sample size. 

We remodelled our gold sample with higher-resolution (0.05\arcsec) simulated HST imaging, which reduced the uncertainty in our posteriors but did not reveal any trends in systematic error. Two systems proved more challenging to model at HST resolution, both of which were Einstein zigzags. These are the most extreme kind of compound lens in which images of S2 are split further by S1 into more images. This makes them even more sensitive to the mass profile of S1. In one case, an SIE description of S1 is required to produce a plausible zigzag model, and in the other both an SIS and SIE model of S1 result in a particularly biased inferred $\beta$. Although these systems are exceedingly rare, forecasts of DSPLs from ongoing surveys such as Euclid \citep{collett_population_2015, li_euclid_2025} and the number of zigzags we find within our simulated compound lens sample suggest $\mathcal{O}(10)$ could be discovered across the EWS, which motivates future investigation into these systems.

Finally, we inferred the cosmological constraints from the combined posteriors of $\beta$ across our simulated sample. Using all 74 of our successfully modelled systems failed to produce physically plausible cosmological constraints due to significant bias in some models. By filtering out systems with tiny Einstein radii, overlapping sources, fewer images and undetectable counters, we produced a gold sample of 19 simulated compound lenses which de-biased our inferred cosmology, leaving a $\sim2\sigma$ tension with the truth but with a large credible region within the $w_0-\Omega_{\rm m}$ parameter space. We improved the model accuracy of our gold sample with simulated higher-resolution HST follow-up, allowing us to produce $50\%$ tighter $1\sigma$ constraints on $w_0$ and a cosmology within $\sim 1\sigma$ of the truth. Ongoing and upcoming surveys from Euclid, Roman and Rubin have the potential to find a sample of DSPLs comparable to our simulations, capable of producing constraints on $w$CDM akin to what we have emulated, retaining that DSPLs are a viable and promising complementary probe of cosmology.

\section*{Acknowledgements}

DJB acknowledges support from the Science and Technology Facilities Council (STFC) through grant ST/3104996/1 and the Royal Society (RF\textbackslash{}ERE\textbackslash{}210033), and thanks Coleman Krawczyk for technical support and guidance. This project has received funding from the European Research Council (ERC) under the European Union’s Horizon 2020 research and innovation programme (LensEra: grant agreement No 945536). TEC is funded by the Royal Society through a University Research Fellowship. AV acknowledges support from STFC ST/X00127X/1 and UKRI7619. 
Numerical computations were performed on both the Sciama High Performance Compute (HPC) cluster (which is supported by the ICG, SEPNet and the University of Portsmouth) and the University of Oxford Advanced Research Computing (ARC) facility  (\hyperlink{https://doi.org/10.5281/zenodo.22558}{https://doi.org/10.5281/zenodo.22558}).

\section*{Data Availability}

The data underlying this article are available in the article.



\bibliographystyle{mnras}
\bibliography{main} 




\appendix

\section{Model Outliers}\label{sec:appA}

\begin{figure}
    \centering
    \includegraphics[width=1\linewidth]{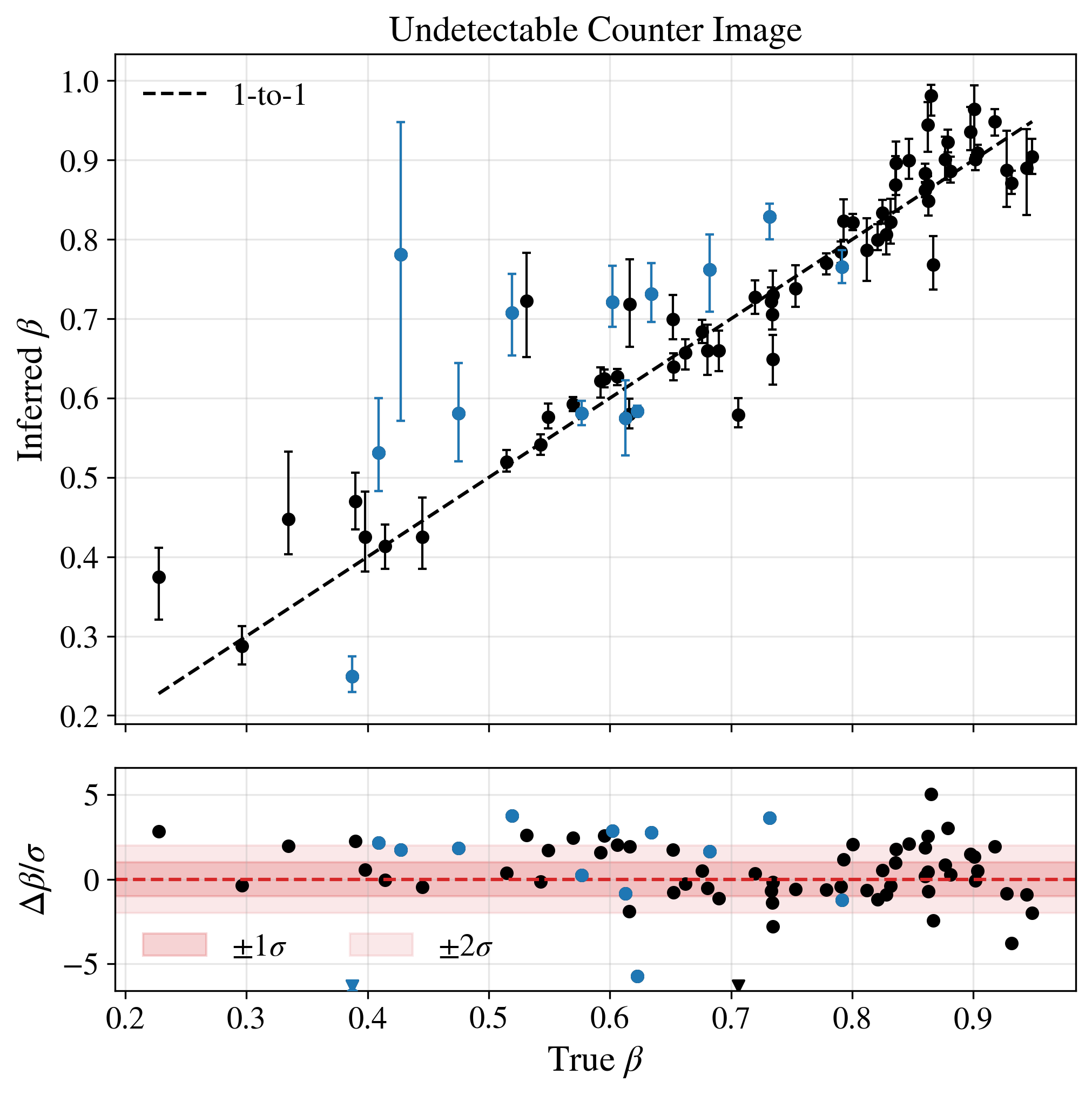}
    \caption{Top and bottom panels show the inferred $\beta$ and normalised residuals against true $\beta$ from models of simulated Euclid compound lenses as described in Figure \ref{fig:beta}. Points in blue highlight systems for which there is no detectable counter image for one source.}
    \label{fig:no_counter}
\end{figure}

In Section \ref{sec:euclid_result}, we discuss the clustering of compound lenses with undetectable and obscured counter images within Figure \ref{fig:beta}. Here, we highlight these systems in blue in Figure \ref{fig:no_counter}. A large fraction of these are distinctly offset from the 1-to-1 line, with most overestimating $\beta$. For those that are overestimates, they also appear to have increased uncertainty in the inferred value. Included in the highlighted systems is the model with the greatest bias in $\beta$ of $6.8\sigma$ and is significantly underestimated compared to the true value. The major bias occurs due to S1 not only lacking a counter image but also having a small Einstein radius of $\theta_{\rm E}^{s1} = 0.364$ arcsec.

We searched for additional qualitative properties of our simulated systems, such as images near the detectability or overlapping thresholds, for similar groupings of models that produced biased $\beta$ posteriors. However, we did not find any evidence beyond the constraints used to produce our gold sample (Section \ref{sec:gold}), which mainly reduced the overall scatter in the recovered $\beta$. Beyond the scope of this work, a larger study into the morphology of lensed image configurations may reveal additional trends.

\section{Lens Models}\label{sec:appB}

\begin{figure*}
    \centering
    \includegraphics[width=0.9\linewidth]{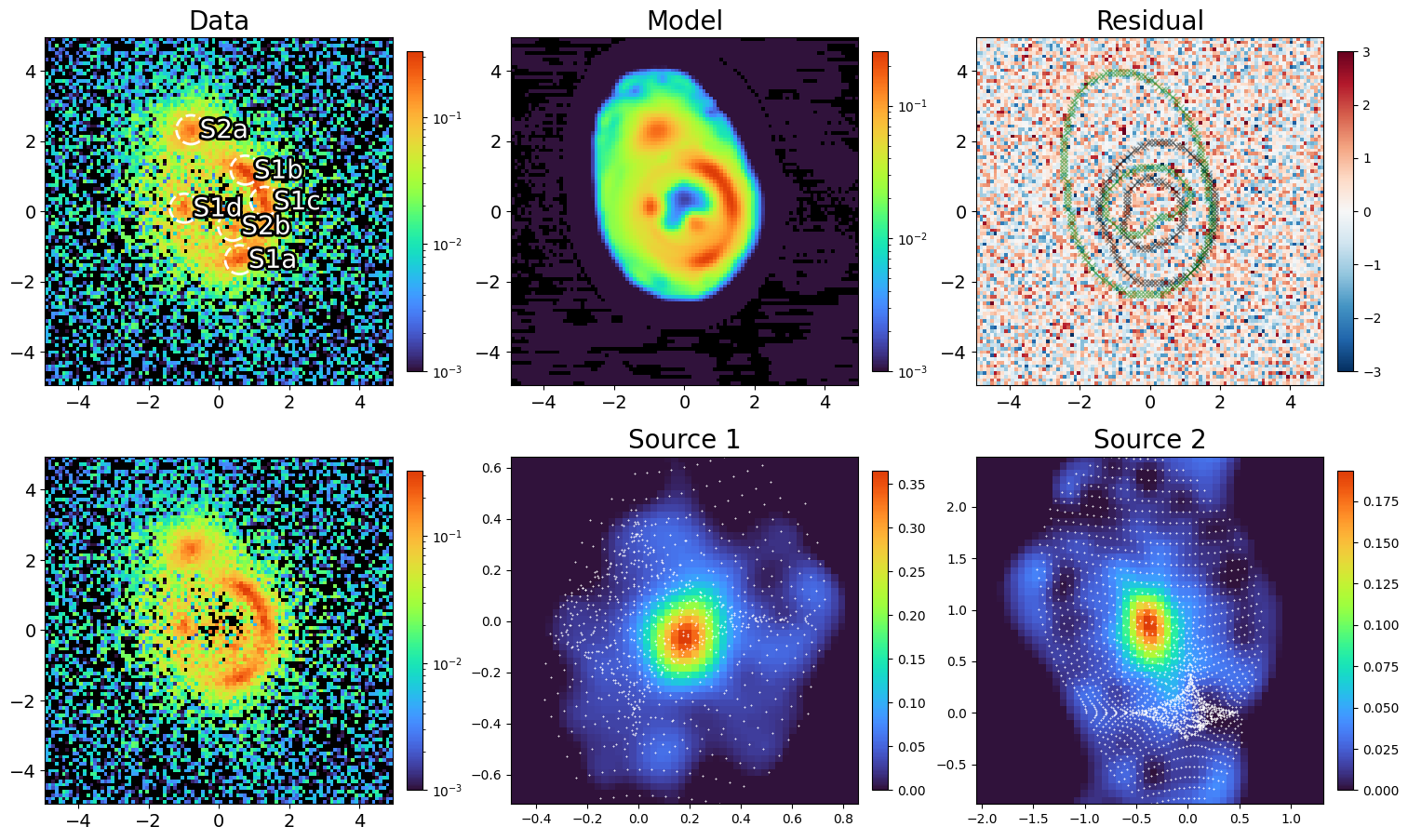}
    \caption{Lens model of one simulated Euclid compound lens produced using \textsc{Herculens} deemed successful. Simulated data is shown in the leftmost column, with the position of each lensed image marked in the top panel by the dashed circles but omitted in the bottom panel for visual clarity. Model reconstruction of the data is shown in the top middle panel, with the normalised residuals between the data and model shown in the top right panel, with source masks outlined in black (S1) and green (S2). Pixelated surface brightness reconstructions of each lensed source are shown in the bottom middle and bottom right panels, where over-plotted white points represent the location of each pixel in the source mask ray-traced from the image plane to the source plane. }
    \label{fig:success_euc}
\end{figure*}

\begin{figure*}
    \centering
    \includegraphics[width=0.9\linewidth]{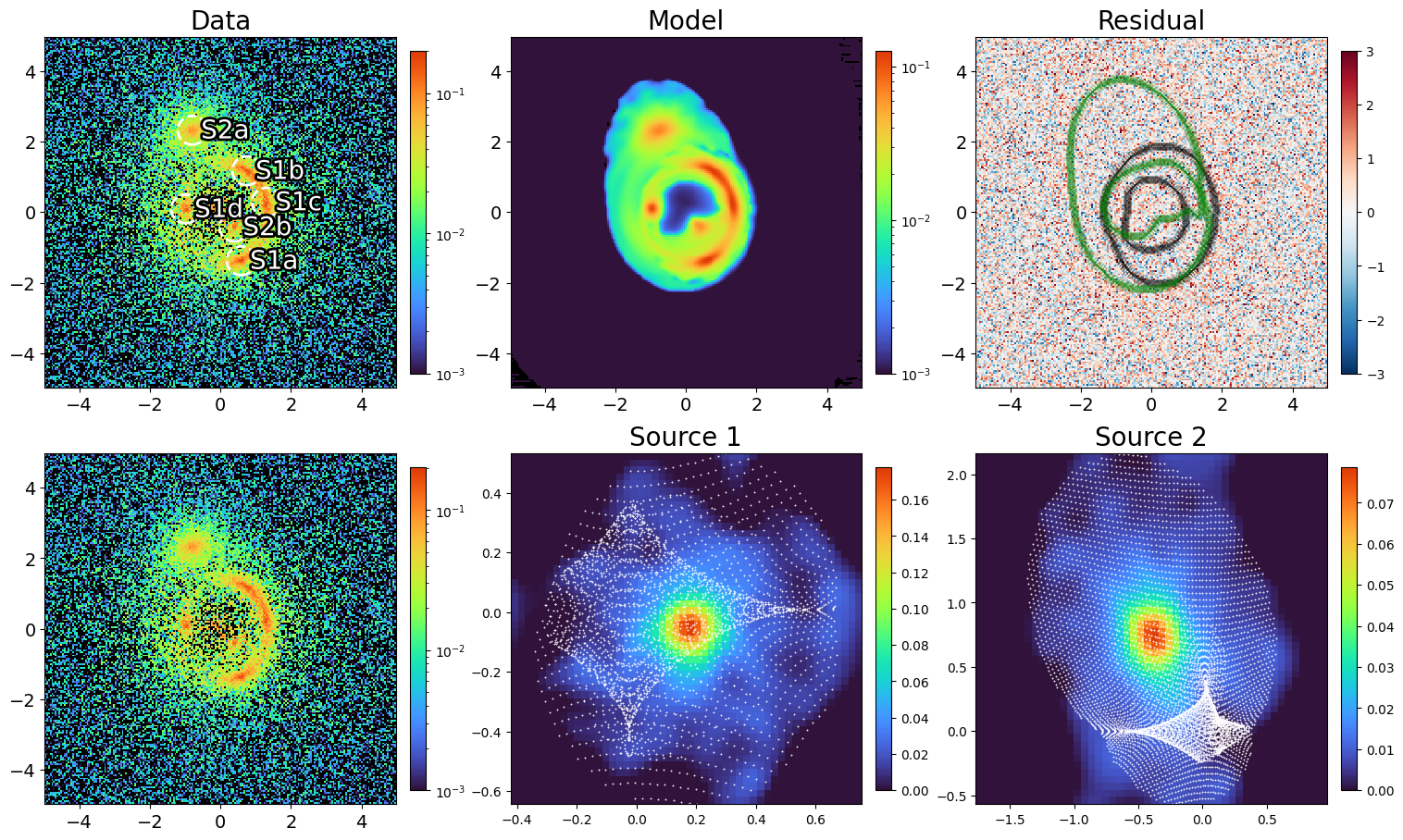}
    \caption{Lens model of the same simulated compound lens as the previous Figure, with Hubble Space Telescope imaging produced using \textsc{Herculens} deemed successful. Panels are as described in the previous Figure.}
    \label{fig:success_hst}
\end{figure*}

\begin{figure*}
    \centering
    \includegraphics[width=0.9\linewidth]{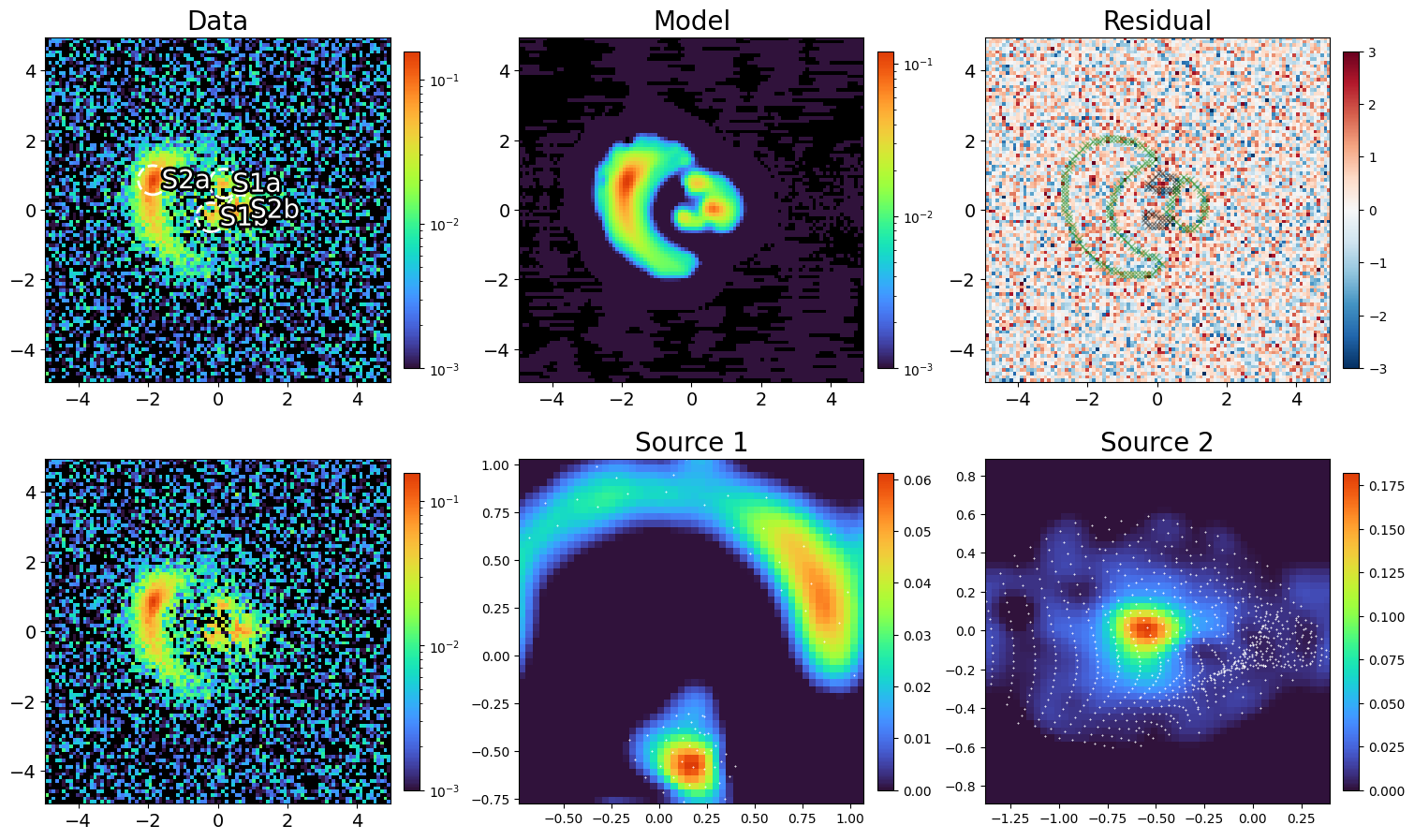}
    \caption{Lens model of one simulated Euclid compound lens produced using \textsc{Herculens} deemed a failure. Panels are as described in the previous Figures. The origin of failure can be seen in the reconstruction of S1 (bottom middle), which has clearly not been traced back to a single source.}
    \label{fig:fail}
\end{figure*}

Figures \ref{fig:success_euc}, \ref{fig:success_hst} and \ref{fig:fail} display lens model summaries generated from the median posterior values of each lens model produced with \textsc{Herculens}. Figure \ref{fig:success_euc} is an example of a successfully modelled Euclid simulation, where residuals are close to the noise level, and the pixelated source light profiles clearly produce focused reconstructions. Figure \ref{fig:success_hst} shows the equivalent model for the simulated HST observations, in agreement with Figure \ref{fig:success_euc}. Some residuals at the core of the lensed images are present in both models, but are more noticeable in Figure \ref{fig:success_hst} due to the greater resolution. These are likely a result of insufficient supersampling or resolution in the pixelated source reconstructions but do not bias the recovered mass-model posteriors.

The failed model shown in Figure \ref{fig:fail} is a typical example of the sampler failing to converge on a focused solution for a source reconstruction, in this case S1. Other failure modes often included very poor residuals and independent HMC chains in disagreement. Many of these lens model failures could have been rectified with manual adjustments to conjugate points, source masks and optimiser iterations. However, this would introduce a large amount of human intervention time for this large, varied sample of compound lens simulations. Hence, we opted for the approach now being taken to automate the strong lens modelling in Euclid DR1 (Euclid Collaboration in prep.). Nonetheless, we acknowledge that for precise cosmological analyses with real DSPLs, a more hands-on approach will be required for each system, as demonstrated for our gold HST sample (Section \ref{sec:gold}).

\bsp	
\label{lastpage}
\end{document}